\documentclass[11pt]{article}
\usepackage{arxiv}
\title{Does High Signal-to-Noise Ratio Identify Better Training Data for Seismic Deep Learning?}
\author[1]{Ziye Yu\thanks{Corresponding author: \href{mailto:yuziye@cea-igp.ac.cn}{yuziye@cea-igp.ac.cn}. ORCID: \href{https://orcid.org/0000-0002-1720-3811}{0000-0002-1720-3811}.}}
\author[2,3]{Xin Liu}
\author[4]{Yuqi Cai}
\affil[1]{Institute of Geophysics, China Earthquake Administration, Beijing 100081, China}
\affil[2]{Laboratory of Seismology and Physics of Earth's Interior, School of Earth and Space Sciences, University of Science and Technology of China, Hefei 230026, China}
\affil[3]{Institute of Advanced Technology, University of Science and Technology of China, Hefei 230088, China}
\affil[4]{University of Chinese Academy of Sciences, Beijing 100029, China}
\date{}
\hypersetup{pdftitle={Does High Signal-to-Noise Ratio Identify Better Training Data for Seismic Deep Learning?},pdfauthor={Ziye Yu, Xin Liu, Yuqi Cai}}
\begin{document}
\maketitle
\begin{abstract}
Hard signal-to-noise ratio (SNR) screening removes weak seismic records, but does retaining clearer waveforms improve learning enough to compensate for the discarded data? We test this question in phase picking and ambient-noise dispersion, first matching training counts and then restoring the full eligible dispersion pool at fixed compute. Phase-picking screening gave no consistent unfiltered-test benefit under fine-tuning or random initialization, including a control matched on distance and phase composition. Dispersion screening increased error at the original five-epoch endpoint but improved on an equally sized unscreened set after longer training. Retaining all 29,788 eligible fit paths instead of 9,929 randomly sampled paths reduced unfiltered-test error by 3.0\% across three seeds at the same update budget. Full-available and strict-SNR training then had similar mean errors, 0.0405 and 0.0406 km s$^{-1}$, with mixed seed-paired ordering; strict training retained a 9.9\% advantage on the SNR-selected test. In two-day monitoring, SNR regulated pick streams but event recovery depended on its measurement definition. High SNR did not reliably identify better training data: gains depended on the task, training schedule, and test domain, and fixed-count comparisons omitted the benefit of retaining additional valid examples.
\end{abstract}

\section*{Key points}
\begin{itemize}

\item Signal-to-noise screening did not consistently improve held-out performance and sometimes reduced it
\item Distance-matched phase-picking training likewise found no extra broad-test benefit from the hard gate
\item Dispersion screening and full-data training had similar mean errors on the unfiltered test
\end{itemize}

\section*{Plain Language Summary}
Seismic models learn from records collected under many conditions. Researchers often discard weak signals because clearer waveforms appear more trustworthy. Yet signal-to-noise ratio measures how strongly a signal stands out from background noise, not how much an accurately labeled example helps a model learn. Screening also leaves fewer examples for training.

We trained models to identify earthquake arrivals and estimate surface-wave speeds. First, we compared screened and unscreened sets with equal numbers of examples. Screening did not consistently improve arrival predictions, and its effect on surface-wave predictions changed with training duration. We then used all available surface-wave training paths, about three times as many, without increasing the number of model updates. Error fell by about 3\% relative to random subsampling. Full-data and screened training had very similar average errors on the unfiltered test, although screened training remained better on a test containing only high-contrast records.

Clearer records therefore do not guarantee better predictions. Screening should be judged against both an equally sized alternative and the full available training set. Its benefits must be weighed against the valid examples it removes, using tests that represent the intended application.

\section*{Keywords}
data-centric machine learning; signal-to-noise ratio; seismic deep learning; training-data curation; selection bias; earthquake monitoring

\section{Introduction}

Does a clearer seismic waveform make a more useful training example? A hard signal-to-noise ratio (SNR) gate assumes that removing low-contrast records will improve the predictions learned from the remaining data. Low SNR can indicate a poor measurement, but it can also describe a valid weak observation. Data-curation choices require scrutiny because removing records is not equivalent to correcting their labels \cite{torralba2011datasetbias,sambasivan2021datacascades}. The scientific question is whether SNR screening improves held-out predictions, and whether any benefit compensates for the valid examples discarded.

SNR is well established as a waveform descriptor in traditional triggers, picker-quality measures, waveform processing, ambient-noise analysis, and machine-learning data sets \cite{allen1978,baer1987,withers1998,boore2005processing,bensen2007ambient,mousavi2019stead,michelini2021instance,zhao2023diting,woollam2022seisbench,park2024hyperparameter}. Some learning workflows also use a scalar SNR cutoff to delete training records \cite{wang2023livestream,ni2023pnw}. Once used this way, SNR becomes an assumption about training value. The score depends on component choice, time windows, and amplitude definition, whereas a record's value for learning depends on its effect on predictions. This differs from methods that estimate label error directly \cite{northcutt2021confident}. Previous work has noted that SNR screening can remove genuine weak arrivals \cite{myklebust2024regional}. Its predictive benefit must therefore be tested directly, with training-set size and evaluation records specified separately: a higher score on an easier selected test does not establish that screening the training data helped.

We first control sample count by comparing each SNR-selected set with an equally sized sample from the full eligible pool. Both models are evaluated on the unfiltered test set and on a test set selected by the same SNR rule. This comparison isolates selection at a fixed data count, but deliberately removes any advantage of retaining more valid records. We therefore add a dispersion model trained on all eligible fit paths at the same number of parameter updates. A further phase-picking control asks whether strict screening adds predictive value when distance bins and phase composition are matched. Separate distance and path-length summaries describe the retained training records.

SNR has a different possible role after inference. A phase picker may produce more candidates than an associator can use efficiently \cite{ross2018gpd,zhu2019phasenet,mousavi2020eqtransformer,zhang2019real}. Here an SNR gate can regulate the pick stream, and its value is determined by event recovery rather than by its usefulness for model training. The operating point may also depend on phase and component because S waves are commonly stronger on horizontal motion and a pre-S noise window can contain P-wave coda \cite{withers1998}. We use model probability as an operational reference, not as a universal quality scale \cite{park2024consistent,puentehuerta2025associators}.

We test the training question in manual-label phase picking and ambient-noise dispersion, using fixed training budgets and cross-evaluation between unfiltered and SNR-selected tests. Phase-picking models are trained from both a released checkpoint and random initialization; extended-training sensitivities test whether the fixed-endpoint conclusions persist. The full-available dispersion condition then tests the additional consequence of reducing the number of distinct training paths. A two-day continuous-association experiment provides a separate comparison of SNR as an operational control. Our focus is the measured gain or loss in predictive performance caused by training-data screening.

\section{Data and Methods}

\subsection{Matched-budget curation framework}

Let $D$ denote an eligible training pool, $q(x)$ a scalar quality proxy, and $C_{\tau}=\{x\in D:q(x)\geq\tau\}$ the candidate pool passing a hard gate. At a common budget $B$, seed $k$ supplied a final selected set $S_{\tau}^{(k)}\subseteq C_{\tau}$ and a comparator $R_{\tau}^{(k)}\subseteq D$, with $|S_{\tau}^{(k)}|=|R_{\tau}^{(k)}|=B$. Phase-picking comparators additionally matched P-only, P+S, and S-only record composition. With a task utility $U$ defined so that larger values are better, the broad-domain and selected-domain curation effects are

\begin{equation}
\Delta_{\mathrm{broad}}=\frac{1}{K}\sum_{k=1}^{K}\left[U\left(f_{S_{\tau}^{(k)}},T\right)-U\left(f_{R_{\tau}^{(k)}},T\right)\right]
\label{eq:broad}
\end{equation}

and

\begin{equation}
\Delta_{\mathrm{selected}}=\frac{1}{K}\sum_{k=1}^{K}\left[U\left(f_{S_{\tau}^{(k)}},T_{\tau}\right)-U\left(f_{R_{\tau}^{(k)}},T_{\tau}\right)\right],
\label{eq:selected}
\end{equation}

where $T$ is the unfiltered held-out set, $T_{\tau}$ is the corresponding SNR-selected test set, and $K=3$ seed-indexed training runs. Comparator sampling also varied by seed when the eligible pool exceeded the common budget. For phase picking, P and S true-positive, false-positive, and false-negative counts were pooled before precision, recall, and F1 were computed; dispersion utility was negative mean absolute error (MAE). Positive values support greater training value under the specified schedule and test domain, not an intrinsic ordering of records. A positive $\Delta_{\mathrm{selected}}$ without a positive $\Delta_{\mathrm{broad}}$ shows a benefit confined to the selected test in that comparison; it does not identify a learning mechanism. Distance and path-length summaries characterize the inputs separately from these prediction outcomes.

These fixed-count contrasts do not include the loss of training records caused by screening. A separate dispersion comparison uses all eligible fit paths, the equally sized full-pool sample, and the strict subset at a common update budget. Full available versus matched full assesses the value of additional distinct paths under that budget; strict versus full available measures the combined effect of changing composition and discarding paths. The latter retains the practical data-quantity consequence of the gate.

Figure \ref{fig:framework} illustrates the screening rule and complementary training comparisons with real records from the dispersion experiment. It separates waveform contrast from the question of predictive value. Table \ref{tab:design} summarizes the controls used in each experiment.

\begin{table}[!htbp]
\caption{Controls used to separate data selection from data quantity. Primary learning comparisons match sample counts; the additional dispersion comparison restores all eligible fit paths at a fixed update budget. The monitoring experiment matches output pick budgets and addresses a distinct event-level question.}
\label{tab:design}
\centering
\small
\begin{tabularx}{\textwidth}{
  >{\raggedright\arraybackslash}p{0.16\textwidth}
  >{\raggedright\arraybackslash}X
  >{\raggedright\arraybackslash}X
  >{\raggedright\arraybackslash}X}
\toprule
Setting & Selection under test & Matched comparator & Evaluation target \\
\midrule
Phase picking & Phase-specific record-level SNR gate & Equal record count and P-only/P+S/S-only composition from the full eligible pool & Unfiltered and equivalently filtered manual-label tests \\
Ambient-noise dispersion & Path-level SNR gate & Equal station-pair count from the full finite-SNR pool & Unfiltered and equivalently filtered station-pair tests \\
Dispersion data quantity & Strict subset versus all eligible fit paths & Same update count, batch sizes, and learning-rate schedule; different unique-path counts & Same held-out station-pair tests \\
Continuous association & Vertical-S, Horizontal-S, or probability ranking & Equal P/S pick budgets and identical REAL configuration & Reference arrivals and catalog events \\
\bottomrule
\end{tabularx}
\end{table}

\subsection{Matched-budget phase-picking experiment}

The phase-picking experiment used manually picked CREDIT-X1local records \cite{li2024creditx1local}. Automatic labels were excluded from training targets and evaluation references. Event keys were disjoint between training and test sets, although stations could occur in both. After requiring 100 Hz three-component waveforms and at least one manual Pg, Sg, Pn, or Sn arrival, 61,429 training records and 16,181 test records were eligible.

Phase-picking SNR was defined as $20\log_{10}(\sigma_{\mathrm{signal}}/\sigma_{\mathrm{noise}})$. P-wave signal amplitude was measured on the vertical component, whereas S-wave signal amplitude used horizontal-vector amplitude. Phase-specific thresholds were calibrated on the training split so that P and S threshold-passing label counts were balanced. A record was retained when either a manual P or S label passed its phase-specific gate, and all manual labels in that record remained in the target. The lower gate used P=14 and S=7.272 dB and retained 45,880 records. The strict gate used P=16 and S=16.598 dB and retained 39,718 records. All final conditions used the common 39,718-record budget and identical P-only, P+S, and S-only composition. The OR rule prevented a single failed phase gate from deleting another manual label in the same waveform. Balanced gate-passing counts do not imply balanced target-label counts after whole-record retention (Text S1).

We trained the PnSn v3 architecture \cite{cai2025pnsn} by fine-tuning its released checkpoint and, separately, from random initialization. For each initialization, full-distribution, lower-gate, and strict-gate subsets were sampled with three seeds. The primary evaluation used deterministic unfiltered manual-label test windows. Cross-evaluation applied the fixed training thresholds and record-level OR rule to the manual test data. A predicted phase was correct when it matched a manual arrival within 1.0 s. Original checkpoints were compared at prespecified final training steps rather than selected by validation performance. As a duration sensitivity, we continued the same full and strict trajectories from 2,000 to 4,000 fine-tuning steps and from 10,000 to 20,000 scratch steps, then scored both endpoints on identical test windows. A separate validation-guided rerun withheld events from the original training split, rematched full and strict subsets, and selected one common checkpoint per initialization mode using only pooled P/S F1 on those validation events. The original test events played no role in checkpoint selection (Text S12). At that rerun's fixed 8,000-step fine-tuning endpoint, an additional control matched the strict set's binned source--station distances and phase composition while retaining the pre-gate pool's binned SNR-margin distribution (Text S13). Texts S1, S12, and S13 give the sample composition, optimization settings, and checkpoint checks.

\subsection{Matched-budget ambient-noise dispersion experiment}

The second learning experiment used SeisDispFusion-NCF \cite{yu2026seisdispfusionncf} and DispNet v2.3, a residual one-dimensional convolutional network that predicts phase velocity at 49 periods. Training and test sets were disjoint by station pair but could share individual stations. The finite-SNR training pool contained 33,098 station pairs, and the two SNR tertiles were 3.04 and 6.77 dB. We compared the full distribution, SNR$>$3.04 dB, and SNR$>$6.77 dB at the common strict-gate budget of 11,033 paths. Models were trained from random initialization with three seeds and evaluated on the same 8,292 unfiltered test paths and 2,734 strict-SNR test paths.

Signal root-mean-square amplitude was measured within the surface-wave arrival window predicted from interstation distance and the valid velocity range. Noise amplitude was measured outside that window, and SNR was $20\log_{10}(\mathrm{RMS}_{\mathrm{signal}}/\mathrm{RMS}_{\mathrm{noise}})$. Each station-pair path was one training sample. Where a path contained multiple valid cross-correlation waveforms (NCFs), its eligibility score was the median of their finite SNR values; training randomly drew one NCF per path presentation. The gate therefore selected paths and their complete period-label masks, not every individual waveform presented during training. This definition is specific to the dispersion task, so its numerical thresholds are not physically comparable with those for phase picking. Original models were evaluated after the prespecified fifth epoch, without validation-based checkpoint selection. A separate duration sensitivity withheld 3,310 training paths for independent validation, rematched the remaining full and strict sets at 9,929 paths, and trained three seeds for 15 epochs. It tested epochs 5, 10, and 15 as specified in advance and also evaluated validation-selected checkpoints on the original held-out test. Because the split, budget, and learning-rate schedule changed, this rerun is not a continuation of the original models (Text S12).

To test the data-quantity consequence of screening, we additionally trained three models on all 29,788 eligible fit paths from that same split. Each used 585 parameter updates and 148,935 path presentations, matching the 15-epoch, 9,929-path models. Batch sizes, loss, optimizer, learning-rate schedule, validation subset, and test inputs were unchanged. The full pool was shuffled in continuous cycles, so every path was presented four or five times rather than 15. The common update endpoint was fixed before the new training; the test did not determine when it stopped. Text S14 gives the full protocol.

\begin{figure}[!htbp]
\centering
\includegraphics[width=\textwidth]{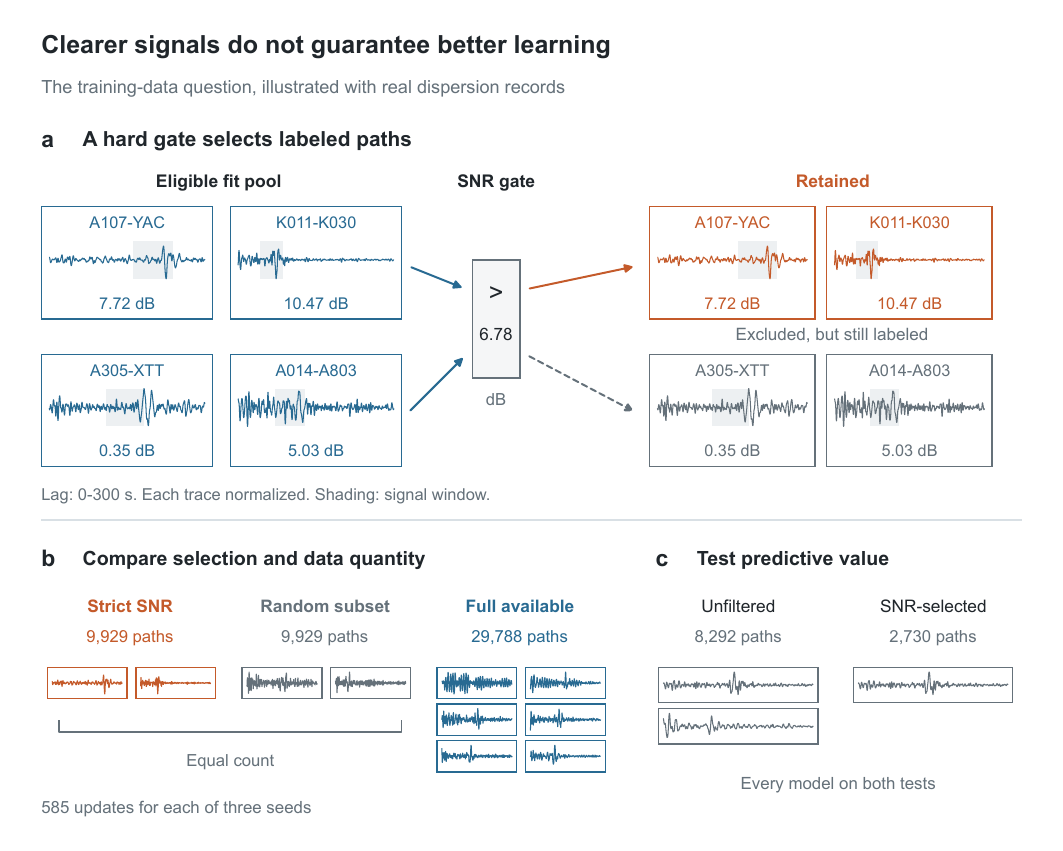}
\caption{Waveform contrast and training value are distinct questions. Real records and counts from the independent-validation dispersion experiment and its full-available extension illustrate the comparison framework (Texts S12 and S14). (a) Observed ambient-noise cross-correlation waveforms from four labeled fit paths. Examples are the nearest single-waveform paths to the 25th and 75th SNR percentiles within the retained and excluded pools. Each trace spans 0--300 s and is normalized by its own maximum absolute amplitude; shading marks the signal window, with the remaining samples defining noise. The strict path-level cutoff is 6.777488 dB, displayed as 6.78 dB. These single-waveform examples have identical waveform and path-median SNR. Screening excludes labeled paths rather than removing noise from their waveforms. (b) The strict and random subsets contain 9,929 paths each; full-available training uses all 29,788 eligible fit paths. Every condition uses 585 updates and 148,935 path presentations for each of three seeds. Thumbnails show actual members of the seed-20260609 manifests, not sample-count estimates. (c) Every model is evaluated on the same 8,292 held-out paths and the 2,730-path SNR-selected subset. Test thumbnails show an actual passing path in both domains and an excluded path only in the unfiltered domain. Test paths are disjoint from fit and validation paths. The diagram explains the controls, not a causal relationship between waveform appearance and prediction error.}
\label{fig:framework}
\end{figure}

\subsection{Continuous association experiment}

Continuous PnSn v3 outputs from 6 July 2019, the Ridgecrest mainshock day, and 13 November 2021, a lower-activity comparison day, were obtained from SeismicX-Cont \cite{yu2026seismicxcont}. Manual S arrivals and preferred network catalog origins served as references and were not generated by the tested inference workflow. Each automatic phase candidate received an amplitude-ratio score from adjacent 2 s pre- and post-pick windows. P-wave SNR was measured on vertical motion. S-wave SNR was measured either on vertical motion (Vertical-S) or with horizontal-vector amplitude (Horizontal-S). A third stream retained candidates by within-phase model-probability rank as an operational reference, not as a universal alternative quality scale.

All streams retained 265,962 P and 265,962 S picks before association. Vertical-S used P$\geq$10.00 and S$\geq$7.451 dB, whereas Horizontal-S used P$\geq$10.00 and S$\geq$8.142 dB. S thresholds were calibrated to the shared phase budgets and held fixed across both days. Every stream was divided into identical 15 min windows and processed with the same REAL configuration. Event recovery was evaluated against 2,340 catalog events using one-to-one matching within 5 s in origin time and 30 km in epicentral distance. Stricter tolerances were tested as sensitivities.

Horizontal-S scoring required valid three-component windows; probability ranking could include candidates without complete component coverage. Their global comparison is therefore an operational stream comparison rather than a score ablation over identical candidates. To isolate component choice, we separately restricted Vertical-S and Horizontal-S rankings to the same candidate IDs with valid three-component data, held P picks fixed, and matched S counts in each association window. Texts S3, S9, and S11 report candidate extraction, threshold calibration, reference provenance, REAL parameters, and matching sensitivities.
\FloatBarrier

\subsection{Uncertainty, reproducibility, and scope}

Learning results are reported across three seed-indexed training runs using means and sample standard deviations. Comparator sampling varied by seed where applicable. The small number of runs supports a stability check but not population-level inference across training archives or architectures. Continuous-monitoring contrasts used paired temporal-block bootstrap intervals: reference-event origin-hour blocks for S-arrival recall and all monitoring-hour blocks for event precision, recall, and F1. Event recovery was also compared with the paired difference in retained S-contributing stations across the complete catalog, without selecting events by association outcome. These intervals quantify stability within the analyzed records and dates, not uncertainty across regions or long-term network conditions. Texts S4--S7 and S9--S11 provide the complete sensitivity analyses.

The analysis environment, dependency specifications, fixed configurations, and reproduction commands are archived with the code described in the Data and code availability section. OpenAI Codex was used for programming assistance, including figure-preparation code, and for language polishing. The authors designed the analyses, verified all code outputs, figures, and reported values, and remain responsible for the manuscript.

\FloatBarrier
\section{Results}

The tests below ask whether screening improves predictions on the same held-out records. Training-set descriptions are reported separately from the performance comparisons and are not used to infer which models should perform better.

\subsection{Training records retained by the gates}

\begin{figure}[!htbp]
\centering
\includegraphics[width=\textwidth]{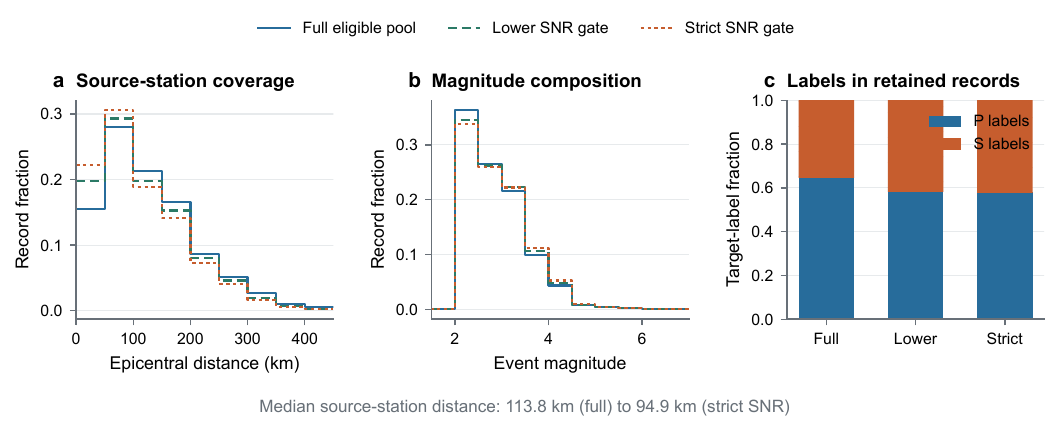}
\caption{Hard phase-specific SNR gates preferentially remove distant phase-picking records before matched-budget sampling. (a) Source--station distance, (b) event magnitude, and (c) P/S target-label composition for the full eligible manual-label pool and the subsets retained by the lower and strict gates. Solid, dashed, and dotted lines distinguish the full, lower-gate, and strict-gate distributions, respectively. Panel (a) shows 0--450 km fractions normalized by all finite-distance records; 0.39\%, 0.21\%, and 0.20\% lie beyond the displayed range for the full, lower-gate, and strict-gate pools. Every label in a record retained by the P-or-S rule remains in its training target. The full and lower-gate pools were subsequently downsampled to the strict-gate budget with identical P-only, P+S, and S-only record composition.}
\label{fig:phase_geometry}
\end{figure}

Before budget matching, SNR selection changed the physical observations available to both learning tasks. The lower and strict phase-picking gates retained 45,880 (74.7\%) and 39,718 (64.7\%) of the 61,429 eligible manual-label records. The strict gate reduced the median source--station distance from 113.8 to 94.9 km, while the magnitude distribution changed little (Figure \ref{fig:phase_geometry}a,b). This contraction remained after budget matching: across seeds, final full-distribution subsets had median distances of 110.2--110.4 km, compared with 94.9 km for the strict subsets (Text S10). Because the record-level OR rule kept every manual label in a retained waveform, this shift was not caused by deleting isolated low-SNR phases; the gate selected a different mixture of records and phase labels (Figure \ref{fig:phase_geometry}c).

\begin{figure}[!htbp]
\centering
\includegraphics[width=\textwidth]{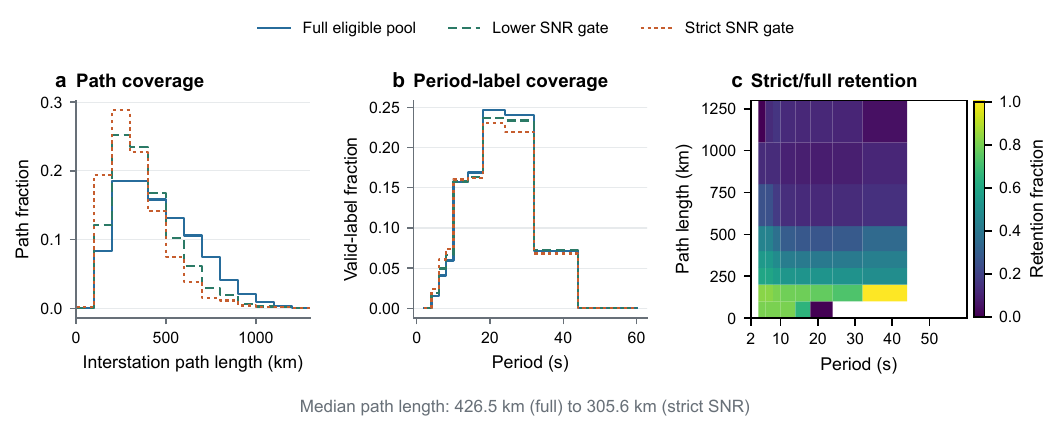}
\caption{Hard dispersion SNR gates preferentially remove long station-pair paths before matched-budget sampling. (a) Complete interstation path-length and (b) valid-period distributions for the full finite-SNR pool and the lower and strict subsets. Solid, dashed, and dotted lines distinguish the full, lower-gate, and strict-gate distributions, respectively. (c) Fraction of full-pool label points retained by the strict gate as a function of path length and period, drawn with the actual unequal bin edges; 60 s is the upper edge of the last period bin, whereas model labels end at 50 s. The full and lower-gate pools were subsequently downsampled to 11,033 paths, matching the strict condition.}
\label{fig:disp_geometry}
\end{figure}

For ambient-noise dispersion, the strict gate retained 11,033 (33.3\%) of 33,098 finite-SNR station pairs and reduced the median path length from 426.5 to 305.6 km (Figure \ref{fig:disp_geometry}a). Its overall period distribution changed modestly, but long paths were depleted across much of the period range (Figure \ref{fig:disp_geometry}b,c). Final full-distribution training subsets had median path lengths of 422.9--424.8 km across seeds, whereas the strict subset remained at 305.6 km (Text S10). In both tasks, the high-SNR pool and the resulting matched training sets represented shorter propagation distances.

\subsection{High-SNR records showed no consistent phase-picking benefit at the original endpoint}

\begin{figure}[!htbp]
\centering
\includegraphics[width=\textwidth]{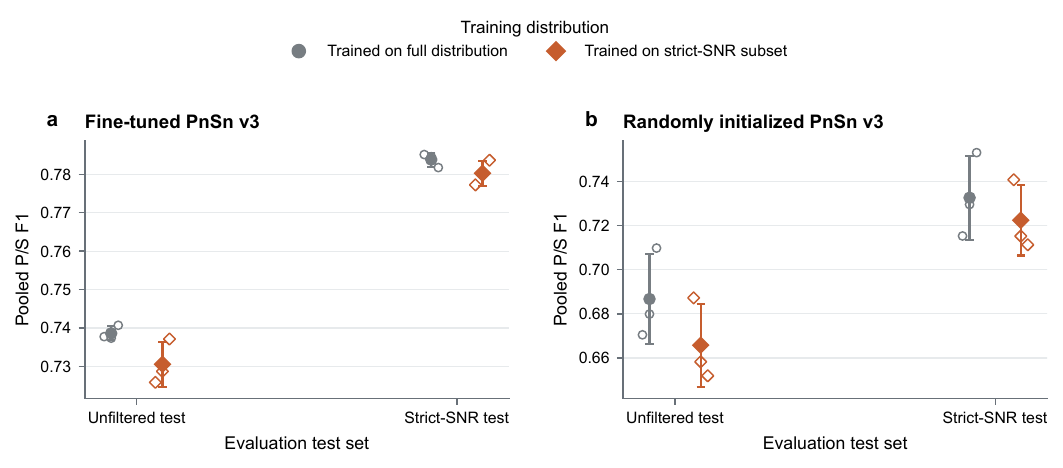}
\caption{At prespecified training endpoints, the strict-SNR subset produced no consistent phase-picking F1 advantage across three runs at a matched 39,718-record budget. (a) Fine-tuned and (b) randomly initialized PnSn v3 models were trained on the full manual-label distribution (gray circles) or the strict-SNR subset (orange diamonds) and evaluated on unfiltered and strict-SNR tests. P and S counts were pooled before F1 was computed. Open symbols show individual runs; filled symbols and error bars show means and sample standard deviations. Both training conditions have identical P-only, P+S, and S-only record composition.}
\label{fig:phase_performance}
\end{figure}

At equal training size and phase composition, the cleaner-looking records showed no stable phase-picking advantage (Figure \ref{fig:phase_performance}). On the unfiltered manual-label test, strict-gate training reduced mean pooled P/S F1 by 0.008 after fine-tuning and by 0.021 after random initialization relative to matched full-distribution training. All three fine-tuning differences were negative; among scratch runs, one increased by 0.017 and two decreased by 0.022 and 0.058. The mean difference was negative under both initialization conditions, but the scratch result was not uniform across seeds.

All models scored higher on the strict-SNR test, which differs from the unfiltered test in both waveform contrast and record composition. Strict-gate training again showed no consistent advantage: mean pooled P/S F1 was lower by 0.004 after fine-tuning and 0.010 after random initialization, although one seed in each initialization condition was slightly positive. Milder-gate differences were smaller and mixed across initialization and test scope (Table S7). Higher scores on the selected test did not establish that selecting the training data by the same rule added value.

\subsection{Strict-SNR paths increased dispersion error at the original endpoint}

\begin{figure}[!htbp]
\centering
\includegraphics[width=\textwidth]{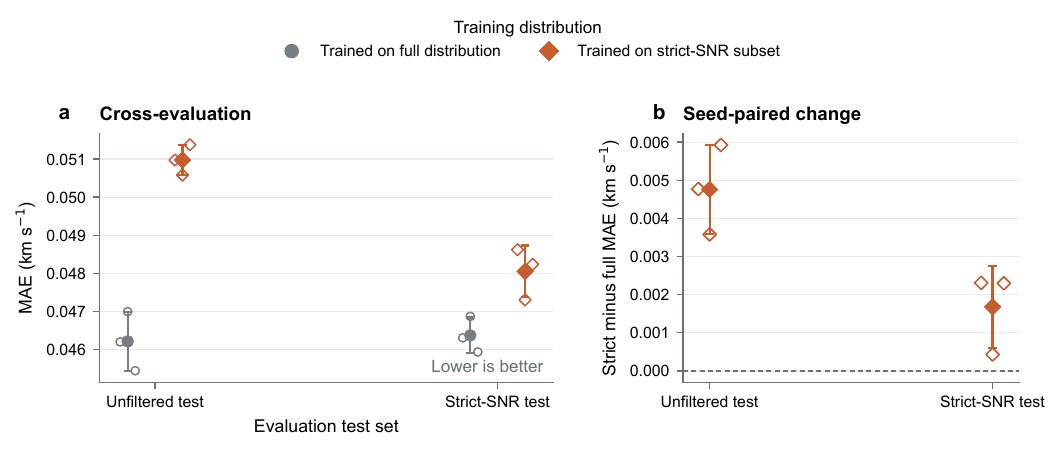}
\caption{At the prespecified fifth epoch, training on strict-SNR station pairs increased dispersion error across three runs at a matched 11,033-path budget. (a) Mean absolute error (MAE) for models trained from random initialization on matched full-distribution (gray circles) or strict-SNR subsets (orange diamonds) and evaluated on unfiltered and strict-SNR tests. (b) Seed-paired strict-SNR-minus-full MAE; positive values indicate larger error after strict-SNR curation. Open symbols show individual runs; filled symbols and error bars show means and sample standard deviations.}
\label{fig:disp_performance}
\end{figure}

Ambient-noise dispersion provided an independent test with different inputs, labels, and loss function. At the prespecified fifth epoch, strict-SNR training increased mean absolute error by 0.0048 km s$^{-1}$ on the unfiltered test and by 0.0017 km s$^{-1}$ on the strict-SNR test (Figure \ref{fig:disp_performance}). Each seed gave the same ordering.

\subsection{Longer training changed the matched-budget ordering}

The phase-picking runs were continued from their exact original checkpoints. At 4,000 fine-tuning steps, the mean strict-minus-full pooled P/S F1 was -0.0025 on the unfiltered test but +0.0026 on the strict-SNR test; all three seed pairs favored strict training on the latter. At 20,000 scratch steps, the corresponding differences were -0.0015 and +0.0094. Scratch differences on the unfiltered test still had mixed signs, and the third strict-test pair was effectively tied (Table S17). These fixed-endpoint continuations weakened the original broad-test disadvantage and exposed a modest selected-test benefit; separate P- and S-wave scores did not always move together (Text S12).

The validation-guided rerun tested whether this ordering survived checkpoint choice independent of the final test. With 35,661 records per condition and the same P-only/P+S/S-only composition, the common validation-selected checkpoints were 8,000 fine-tuning and 24,000 scratch steps. Strict-minus-full pooled F1 averaged -0.0027 and -0.0100, respectively, on the unfiltered holdout; on the strict-SNR holdout it averaged +0.0012 and -0.0060. The scratch differences changed sign across seeds. The fine-tuning validation score was still rising at the 8,000-step compute limit, while the scratch score fell at 28,000 after peaking at 24,000; neither mode satisfied the prespecified plateau rule (Table S19). Thus the selected-test advantage seen at a fixed continuation endpoint was not stable across initialization and validation-guided checkpoint choice, and these runs do not establish full convergence.

In the distance-matched fine-tuning control, strict-SNR training gave no consistent gain over pre-gate sampling on the unfiltered holdout: mean pooled F1 was 0.7395 versus 0.7392, with mixed seed-paired differences. On the similarly filtered holdout, strict training scored 0.7919 versus 0.7873. The control matched phase composition and binned distance, but not other source, station, or waveform attributes (Text S13; Table S20).

The dispersion duration sensitivity used a new fit--validation split rather than continuing the original checkpoints. In that matched 9,929-path rerun, full-distribution training had lower unfiltered-test MAE at epoch 5 in all three pairs. By the common, prespecified epoch 15, strict-SNR training had lower MAE in all three: 0.0406 versus 0.0418 km s$^{-1}$ on the unfiltered test and 0.0373 versus 0.0426 km s$^{-1}$ on the strict-SNR test, averaged across seeds (Table S18). Across the six models, mean validation MAE at epoch 15 was within 0.00004 km s$^{-1}$ of its lowest value over the 15-epoch run (Text S12). The third unfiltered pair differed by only 0.0001 km s$^{-1}$, and this validation check does not prove convergence or establish a unique stopping epoch. Because the validation split changed the training budget and schedule, these values test sensitivity to optimization duration rather than replace the original five-epoch estimates.

The two learning tasks use different SNR definitions, models, inputs, and targets. SNR screening did not yield consistent predictive gains: several comparisons favored unscreened training, while the dispersion ordering reversed under longer training. The predictive value of the gate therefore cannot be inferred from waveform contrast alone.

\subsection{Dispersion screening versus full-available training}

Restoring all 29,788 eligible fit paths reduced unfiltered-test MAE from 0.04177 for the 9,929-path random sample to 0.04052 km s$^{-1}$, a 3.0\% reduction at the same 585 updates. All three seeds improved; error also fell by 2.7\% on the strict-SNR test (Text S14; Table S21). More distinct training paths therefore improved on random subsampling without extra parameter updates.

Strict-SNR training had no mean MAE advantage over full-available training on the unfiltered test: 0.04060 versus 0.04052 km s$^{-1}$, with mixed seed-paired ordering. These close means do not establish statistical equivalence. Strict training did retain lower error on its selected test, 0.03730 versus 0.04142 km s$^{-1}$, a 9.9\% reduction in the three-seed mean. A gain relative to random subsampling therefore did not imply a gain over the full eligible pool on the unfiltered test.

\subsection{SNR serves a different role in continuous monitoring}

\begin{figure}[!htbp]
\centering
\includegraphics[width=\textwidth]{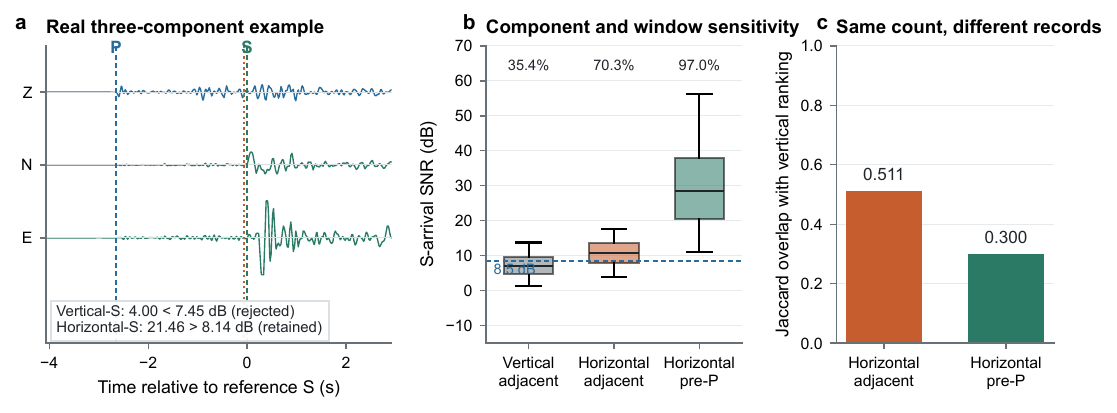}
\caption{The S-arrival SNR scale and the identities of retained arrivals depend strongly on component and noise-window definition. (a) Real three-component waveform for an automatic S candidate that is rejected by the Vertical-S definition and retained by the Horizontal-S definition. (b) Distributions for 8,374 held-out manual S arrivals under vertical-adjacent, horizontal-adjacent, and horizontal pre-P definitions; boxes show the interquartile range, whiskers show the 5th--95th percentiles, and percentages report the fraction above a common 8.5 dB threshold. The pre-P window uses a reference P time and is a nondeployable mechanism diagnostic. (c) Jaccard overlap with the vertical ranking after fixing every definition to the 2,965-arrival budget of the vertical-adjacent gate.}
\label{fig:component_sensitivity}
\end{figure}

After inference, SNR no longer selects what a model learns; it selects which picks reach the associator. The component used to measure S-wave energy then matters. The same candidate S arrival failed the Vertical-S threshold but passed the Horizontal-S threshold (Figure \ref{fig:component_sensitivity}a). Across 8,374 held-out manual S arrivals, horizontal-adjacent SNR was a median 3.449 dB higher than vertical-adjacent SNR. A reference-timed pre-P noise window raised it by a further 17.424 dB, confirming the influence of P-wave coda but requiring information unavailable in routine deployment (Figure \ref{fig:component_sensitivity}b; Table S8). At the same retained count, the vertical ranking had Jaccard overlaps of only 0.511 and 0.300 with the horizontal-adjacent and horizontal-pre-P rankings (Figure \ref{fig:component_sensitivity}c).

\begin{figure}[!htbp]
\centering
\includegraphics[width=\textwidth]{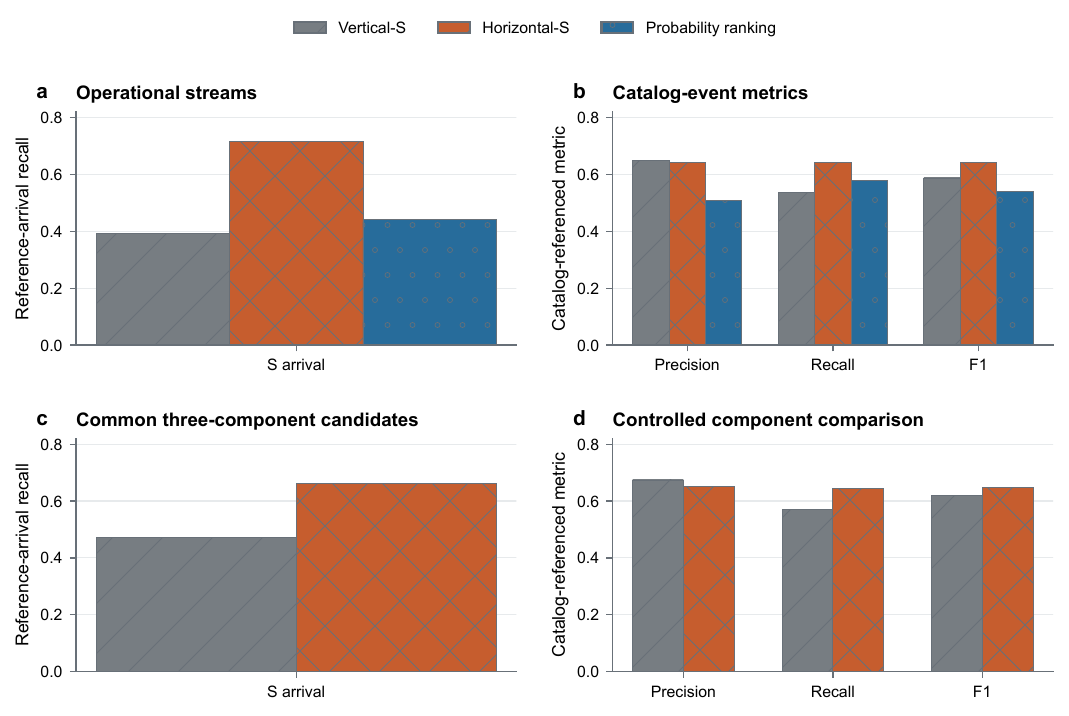}
\caption{At equal P/S pick budgets, the S-wave component definition changes arrival and catalog-event recovery in the two-day continuous experiment. (a, b) Operational Vertical-S (gray diagonal hatch), Horizontal-S (orange cross-hatch), and within-phase probability-ranking streams (blue dotted hatch), each retaining 265,962 P and 265,962 S picks before the same REAL workflow. Horizontal-S and probability ranking have different global candidate eligibility. (c, d) Controlled component comparison restricted to the same valid three-component S-candidate IDs, with identical P picks and S counts matched in each 15 min association window. Metrics are reference-arrival recall and catalog-referenced event precision, recall, and F1.}
\label{fig:monitoring_metrics}
\end{figure}

At the same global phase budget, Horizontal-S recovered 71.8\% of reference S arrivals, compared with 39.4\% for Vertical-S (paired difference 0.324, 95\% confidence interval [0.314, 0.335]; Figure \ref{fig:monitoring_metrics}a). The additional S support propagated to association: catalog-event recall rose from 0.536 to 0.643 (difference 0.106 [0.091, 0.123]), and event F1 also increased, while the precision difference was not resolved (Figure \ref{fig:monitoring_metrics}b). The two streams carried the same numbers of P and S picks, but not the same event information.

Candidate availability did not explain this difference. When both definitions ranked the same three-component candidates, with identical P picks and the same S count in every association window, Horizontal-S increased S-arrival recall from 0.473 to 0.662 and catalog-event recall from 0.571 to 0.645 (Figure \ref{fig:monitoring_metrics}c,d; Table S15).

\begin{figure}[!htbp]
\centering
\includegraphics[width=\textwidth]{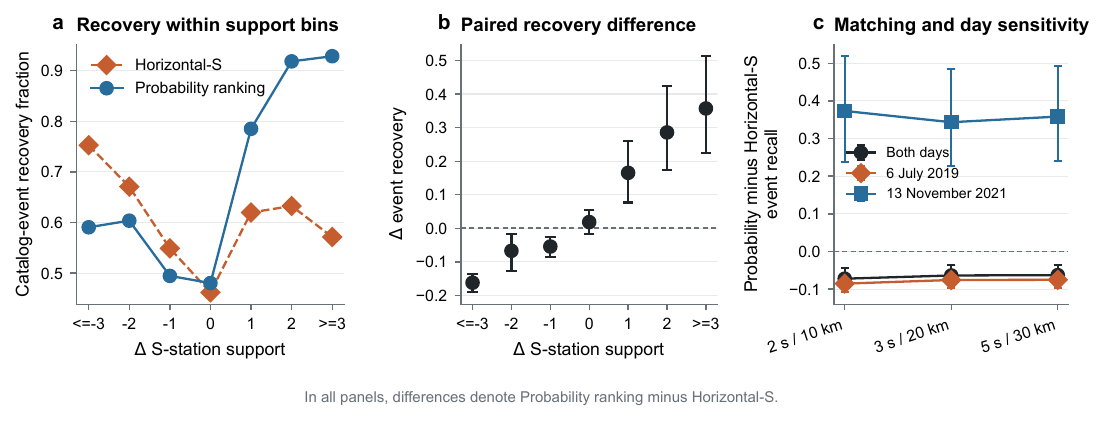}
\caption{Catalog-event recovery covaries with retained S-station support, while the operational ordering varies by monitoring day. (a) Recovery fractions and (b) paired recovery differences for all 2,340 catalog events binned by the corresponding difference in retained S-contributing stations. Bin counts from left to right are 999, 328, 368, 433, 121, 49, and 42 events, and error bars are 95\% origin-hour block-bootstrap confidence intervals. (c) Event-recall differences under three event-matching tolerances for both days (black circles), the 2,273-event mainshock day (orange diamonds), and the 67-event lower-activity day (blue squares). Differences in all panels are Probability ranking minus Horizontal-S, so negative values favor Horizontal-S. The global streams have equal P/S output budgets but different candidate eligibility.}
\label{fig:support_recovery}
\end{figure}

Probability ranking provides an operational reference, not a universal benchmark. Across all 2,340 catalog events, the stream retaining more S-contributing stations also tended to recover more events (Figure \ref{fig:support_recovery}a,b). Horizontal-S recovered 0.063 more of the catalog than probability ranking under the global budget (95\% confidence interval [0.035, 0.085]). This ordering was insensitive to event-matching tolerance in the combined data and on the mainshock day, but it reversed on the 67-event lower-activity day (Figure \ref{fig:support_recovery}c). No single ranking was validated across both monitoring regimes.

\FloatBarrier
\section{Discussion}

\subsection{Waveform clarity is not a measure of learning contribution}

Whether a record is easy to recognize and whether it helps a model interpret other records are different questions. SNR measures contrast under a specified component and window definition; it does not directly measure label reliability or a record's contribution to learning. A weak arrival can have a reliable manual label, so excluding it need not remove erroneous supervision. Conversely, high contrast does not establish that an example adds information beyond the records already retained. Figure \ref{fig:framework} makes this distinction explicit: the gate selects labeled observations, while their training value is tested through predictions on held-out records.

The shorter distances and paths retained by screening document which physical observations the gate selects. They are evidence of structured selection, not a sufficient explanation of the learning outcome. The distance-matched phase-picking control directly tested the added value of the gate at the same record count, source--station distance-bin counts, and phase composition, while preserving the pre-gate pool's coarse SNR-margin distribution. Mean pooled F1 on the unfiltered holdout was 0.7395 for strict training and 0.7392 for the comparator, with mixed seed-paired ordering; a small, consistent gain remained on the SNR-selected test (Text S13; Table S20). Thus the absence of a consistent broad-test benefit persisted against this geometrically matched alternative, rather than depending only on a comparison with a broader-distance training set.

Matching distance does not match the waveform structure available at that distance. One possible explanation is that clearer records can repeat already represented signal patterns, whereas some correctly labeled weak records add variations that are useful on other observations. Screening could then reduce exposure to examples requiring temporal structure, cross-component relationships, or dispersion continuity rather than strong local amplitude contrast. This is a hypothesis about learning contribution, not a mechanism identified by our experiments: the controls do not measure learned representations or the marginal value of individual records. Nor does this hypothesis imply that noise itself is beneficial or that low-SNR examples are intrinsically more useful.

The observed outcomes support that distinction without requiring a common response from both tasks. Strict screening gave no consistent unfiltered-test phase-picking gain and scored lower under several protocols, whereas dispersion screening helped after longer training, especially on the selected test. Training value must therefore be assessed relative to the other available examples, the learning protocol, and the intended predictions, rather than assigned from waveform contrast alone.

\subsection{What the matched-budget controls isolate}

The three training conditions in Figure \ref{fig:framework} answer complementary questions. Strict versus equally sized random training tests selection at a fixed number of distinct examples. Random versus full-available training tests the benefit of retaining additional eligible observations under the same update budget. Strict versus full available tests the practical consequence of applying the gate, combining its change in sample composition with its reduction in sample count. The equal-count random subset is therefore a valid control for selection, but it cannot alone establish that discarding the remaining data improves learning.

The dispersion results illustrate the distinction. Retaining all eligible fit paths improved on random subsampling in every seed, reducing mean unfiltered-test MAE by 3.0\%. The strict subset's unfiltered-test advantage over random subsampling was not observed against full-available training. Strict and full-available means were close, however, with mixed seed-paired ordering and a small reversal of their mean ordering for RMSE (Text S14; Table S21). These results do not establish either equivalence or a consistent full-data advantage over screening.

Viewed from the selection side, about one third of the eligible dispersion paths reached nearly the same mean unfiltered-test MAE as full-available training and a 9.9\% lower mean MAE on the selected test. This is evidence of useful selection under that protocol, although not of computational acceleration: every condition used the same 585 updates, with more repetitions per path in the smaller sets. Evaluating every model on both test domains distinguishes changes caused by training selection from changes caused by evaluating only selected records. Applying the same threshold to training and test does not, however, match all their other physical or statistical properties.

Training duration also changes the answer. The original fixed endpoints showed no stable selected-domain advantage, but longer training favored strict screening in some phase-picking comparisons and reversed the dispersion ordering. Fixed-endpoint comparisons estimate performance under a specified training allowance; validation-based checkpoint selection assesses performance under a specified selection rule. Our phase validation reruns reached their compute limits without meeting the plateau criterion, and the dispersion rerun used a new fit split and schedule (Text S12). The comparisons therefore support protocol-specific conclusions, not a ranking that holds at every stage of optimization. Soft weighting, curricula, and direct label-error detection remain separate questions.

\subsection{Training selection and monitoring control answer different questions}

Training curation asks whether removing records helps a model learn. Monitoring asks which picks should be passed to an associator under a finite processing budget. SNR can be useful for the second purpose even when it is unhelpful for the first. In the continuous experiment, equal pick counts produced different event catalogs because the Vertical-S and Horizontal-S definitions retained different S arrivals and station support.

The component comparison tests the observing workflow rather than introducing a new SNR estimator or a universal Horizontal-S threshold. Changing the component changed which phases supported each event. The numerical operating point also depends on the picker, preprocessing, available components, measurement windows, associator, and event rate. Because the thresholds were calibrated and evaluated on the same two days, they require independent validation before operational use.

Probability ranking has the same limitation as an operating reference. Neural-picker scores inherit the training distribution and need not be calibrated across stations or dates. The reversal between the mainshock and lower-activity days shows that neither ranking can be transferred without testing the event-level trade-off on the continuous stream where it will be used.

\subsection{A reusable audit of proxy-based data curation}

A reusable audit should establish when a proxy score predicts downstream value and what is lost when it is used as a hard cutoff. The present comparisons suggest six steps. (1) Specify the measurement and unit of selection, including components, windows, units, thresholds, and whether a score removes a phase label, a record, or an entire path. (2) State the intended benefit and evaluation target, distinguishing label reliability, measurement usability, sample difficulty, and training contribution. (3) Compare selected, equally sized random, and full-available training sets, matching relevant task strata and reporting distinct-example counts, update budgets, and repetitions separately.

(4) Describe how retention changes physical coverage and test proposed explanations with targeted controls, such as distance matching; a changed histogram alone does not identify a learning mechanism. (5) Specify training and validation-based selection rules before test evaluation, examine duration sensitivity, and evaluate every model on the same intended and gate-selected tests. (6) Assess the consequence at the final decision stage: for training, whether predictive gains justify discarding eligible examples; for monitoring, whether retained picks support the intended event-level outcome.

The audit turns a presumed quality criterion into falsifiable comparisons of selection, data quantity, and target-domain performance. Its matched-budget logic is model-agnostic and dataset-agnostic and can be applied to uncertainty scores, model probabilities, waveform measures, or metadata rules. Transformer architectures and global seismic archives are natural settings for such tests, not settings in which the present numerical ordering can be assumed. The reusable contribution is a way to determine when screening earns its place in a workflow, rather than a replacement scalar score claimed to measure training value everywhere.

\subsection{Limits of inference}

The learning experiments cover two archives, one architecture per task, and three training seeds. Training and test records are separated by event or station pair, but stations can occur in both. The original dispersion loss was still decreasing at epoch 5. The first phase-picking continuation reused its original training trajectories; a further phase sensitivity instead withheld validation events, used a smaller matched budget, and selected a common checkpoint without consulting the original test. The dispersion sensitivity also used an independent validation split, a smaller matched budget, and an extended learning-rate schedule (Text S12). Its original test split remained held out, but the new strict-test membership differed slightly because the cutoff was recalibrated on fit paths. These checks expose training-duration dependence; the phase rerun reached its compute limits without meeting the prespecified validation-plateau rule, so the reported scores are budget-limited comparisons rather than convergence estimates. The additional phase control matches binned distance and SNR margin, not continuous distributions or all source, station, and waveform covariates; its training manifests overlap the strict sets and it was tested only for fine-tuning. There is no corresponding dispersion geometry control, and we have not identified which individual record properties account for the score differences. Replication across two tasks does not establish the same response for other architectures or regions.

The full-available dispersion test fixes updates and total path presentations rather than repetitions per path: the larger set is presented four or five times, versus 15 for the smaller sets. Reference dispersion curves were also estimated separately from each condition's training paths (Text S14). The comparison evaluates complete training workflows under a common update budget, not an isolated effect of record count or the best attainable performance after separately optimizing each condition. Its near-equal unfiltered-test means are not an equivalence result, and its data-quantity finding has not been tested in the phase-picking task.

The monitoring experiment is more limited. It covers two days, one picker, one associator, and one regional catalog, with thresholds calibrated on the records used for evaluation. Catalog incompleteness may also affect catalog-referenced precision differently among streams. The common-candidate comparison isolates the component ranking, but longer continuous records and independent regions are needed before any operating threshold is transferred.

\section{Conclusions}

High waveform contrast does not by itself identify the records that contribute most to learning. In phase picking, strict screening provided no consistent unfiltered-test gain, including in fine-tuning matched on distance and phase composition, and reduced performance under several protocols. In dispersion, screening increased error at the original endpoint but improved on an equally sized random subset after longer training. Restoring all eligible dispersion paths reduced mean unfiltered-test MAE by 3.0\% relative to random subsampling at the same update budget. Strict and full-available training then had close mean errors with mixed seed-paired ordering, while strict training retained a 9.9\% selected-test advantage. These findings establish conditional benefits and costs of screening, not a universal benefit from either higher contrast or lower SNR.

The monitoring experiment addresses a different decision. SNR regulated a continuous pick stream, but event recovery depended on the S-wave measurement definition and day. Utility after inference does not establish that the same measure should be used to discard training records.

A claim that screening improves learning requires comparison with both an equally sized alternative and the full eligible pool under an explicit resource budget. Coverage summaries show what is selected; controlled prediction tests establish the consequences for learning. Our six-step audit makes that distinction testable across models, data sets, and quality proxies. Training value belongs to the relationship between an example, the other available data, and the target task, not to waveform contrast in isolation.

\section*{Acknowledgements}
The authors acknowledge the public data and software resources cited in the Data and code availability section. We thank the developers of Python, NumPy, SciPy, pandas, h5py, Matplotlib, PyTorch, and ObsPy. This work was supported by the China Earthquake Administration under Grant No. DQJB25B43.

\section*{Conflict of Interest}
The authors declare there are no conflicts of interest for this manuscript.

\section*{Data and code availability}
Analysis scripts, checkpoints, seed-fixed manifests, derived tables, figure data, and environment specifications for the original fixed-endpoint experiments are archived in \textit{seismic-snr-filtering-bias} \cite{yu2026seismicsnrfilteringbias} under CC BY 4.0. The archive is available at \url{https://huggingface.co/cangyeone/seismic-snr-filtering-bias}. Third-party raw waveform and dispersion arrays are not redistributed. Continuous waveforms and labels are available from SeismicX-Cont \cite{yu2026seismicxcont}, and dispersion data are available from SeisDispFusion-NCF \cite{yu2026seisdispfusionncf}. CREDIT-X1local phase-picking data are available from the National Earth System Science Data Center \cite{iesdc2023creditx1local}; registration and acceptance of its data-sharing agreement are required. The data set is described by \citet{li2024creditx1local}. Data use follows each source archive's license.

\FloatBarrier
\bibliographystyle{unsrtnat}
\bibliography{references}

\clearpage
\suppressfloats[t]
\setcounter{figure}{0}
\setcounter{table}{0}
\setcounter{equation}{0}
\setcounter{section}{0}
\renewcommand{\theHfigure}{S.\arabic{figure}}
\renewcommand{\theHtable}{S.\arabic{table}}
\renewcommand{\theHequation}{S.\arabic{equation}}
\renewcommand{\theHsection}{S.\arabic{section}}
\renewcommand{\thetable}{S\arabic{table}}
\renewcommand{\thefigure}{S\arabic{figure}}
\renewcommand{\theequation}{S\arabic{equation}}
\graphicspath{{figures/}}
\section*{Supplementary Information}
\noindent\textbf{Contents of this file}

\noindent Text S1. Phase-picking experiment parameters

\noindent Text S2. Dispersion experiment parameters

\noindent Text S3. Continuous association parameters

\noindent Text S4. Paired bootstrap parameters

\noindent Text S5. Additional tabulated uncertainty and association summaries

\noindent Text S6. Event-level retained-geometry robustness and local threshold sensitivity

\noindent Text S7. Vertical-only continuous-filter sensitivity

\noindent Text S8. Label-independent case-event repicking

\noindent Text S9. S-arrival component and noise-window sensitivity

\noindent Text S10. Observation-geometry shifts under training curation

\noindent Text S11. Component and eligibility sensitivity

\noindent Text S12. Extended-training and validation sensitivity

\noindent Text S13. Joint distance and pre-gate SNR-distribution control for phase picking

\noindent Text S14. Full-available dispersion training at a fixed update budget

\subsection*{Text S1. Phase-picking experiment parameters.}
The phase-picking experiment used \path{data/credit-x1.h5} and the train/test event keys in \path{data/creditkeys.npz}. The archive event keys were disjoint between train and test, whereas all 349 test stations also occurred in the training split; the experiment is therefore event disjoint but not station disjoint. After requiring 100 Hz BHE, BHN, and BHZ waveforms and at least one manual Pg, Sg, Pn, or Sn arrival, the train and test splits contained 61,429 and 16,181 station records. The training pool contained 60,318 manual P and 32,856 manual S labels; automatic \texttt{RNN.tagY} labels were excluded from both training targets and evaluation references. This source restriction prevents automatic labels related to a phase picker from determining the reported performance.

Phase-picking SNR was $20\log_{10}(\sigma_\mathrm{signal}/\sigma_\mathrm{noise})$. For P, signal standard deviation used the vertical component in the 0.5 s after the manual P arrival. For S, signal standard deviation used the horizontal-vector amplitude $[(\sigma_E^2+\sigma_N^2)/2]^{1/2}$ in the 1.5 s after the manual S arrival. The noise estimate used the same component convention in a 1.5 s pre-event window ending 2 s before the nearest preceding manual P arrival. For the 1,716 S-only training records without a preceding P, horizontal noise was measured in the first 1.5 s of the waveform. These values are task-specific filtering coordinates whose thresholds define the selected training distribution.

Threshold pairs were calibrated from the training split. The lower gate used P=14 and S=7.272 dB and retained 31,966 threshold-passing labels of each phase; the higher gate used P=16 and S=16.598 dB and retained 27,606 of each. Filtering then used the prespecified record-level OR rule: a waveform was retained if any manual P or S label passed its phase-specific threshold, and all manual P/S labels in that waveform remained in the training target. Candidate pools contained 61,429 full records, 45,880 lower-gate records, and 39,718 higher-gate records, corresponding to 100\%, 74.7\%, and 64.7\% retention. All final conditions used the maximum common 39,718-record budget and identical composition: 11,575 P-only, 27,036 P+S, and 1,107 S-only records. The corresponding retained target-label counts were 38,990--39,052 P and 28,171--28,173 S; these differ from the exactly balanced counts that passed the gates because the OR rule preserves every manual label in a retained waveform.

For each matched training subset, we ran two initialization modes with the BRNNPNSN architecture: fine-tuning from \path{ckpt/pnsn.v3.pt} and training the same architecture from random initialization. Pg and Pn outputs were merged as P, and Sg and Sn outputs were merged as S. Fine-tuning was run for 2,000 iterations, and scratch initialization was run for 10,000 iterations, both with batch size 16, AdamW learning rate $10^{-4}$, and weight decay $10^{-4}$. Training and matched-subset sampling used seeds 20260609, 20260610, and 20260611. The full and lower-gate candidate pools were resampled for each seed to match the higher-gate pool composition; batch draws varied for every model. The prespecified final-step checkpoints were evaluated without validation-based checkpoint selection. The primary evaluation used 10,000 deterministic unfiltered, manual-label test windows per seed. A predicted pick was counted as correct if it matched a manual P or S arrival within 1.0 s. For the reported pooled P/S F1, true-positive, false-positive, and false-negative counts were summed across P and S before precision, recall, and F1 were calculated. Cross-evaluation applied the fixed training thresholds to the manual test records and retained complete records under the same OR rule; every model was evaluated on the same windows and labels within each test scope. Text S12 reports a later duration sensitivity using the same matched training trajectories.

\subsection*{Text S2. Dispersion experiment parameters.}
The dispersion experiment used \path{data/ncf_data/ncf_disp_dataset_with_disp_image.h5}. Each station-pair sample contains one or more ambient-noise cross-correlation waveforms, station coordinates, a 2--50 s period grid, phase-velocity labels, and a validity mask. We trained DispNet v2.3, a residual one-dimensional convolutional network that predicts 49 period-dependent phase velocities from one cross-correlation waveform at a time, from random initialization. The archive training and test sets contained 33,098 and 8,292 distinct station pairs with zero canonical-pair overlap; all 394 stations occurred in both splits. The split was therefore station-pair disjoint but not station disjoint.

NCF SNR was defined from the predicted surface-wave arrival window. Station distance and the valid dispersion-velocity range determined the expected arrival window; signal RMS was measured within this window, noise RMS outside it, and SNR was defined as $20\log_{10}(\mathrm{RMS}_\mathrm{signal}/\mathrm{RMS}_\mathrm{noise})$. One station-pair key was one path-level training sample with a shared period-label mask. We computed SNR for each valid NCF under that key, discarded nonfinite values, and used their median as the path eligibility score. During training, each visit to a selected path drew one of its valid NCFs at random; deterministic evaluation used the first valid NCF in sorted order. Of 33,098 finite-SNR training paths, 2,092 had more than one valid NCF. Among the 11,033 paths above the exact strict cutoff of 6.774182 dB, 511 had multiple NCFs, and 336 of those (3.0\% of strict-selected paths) had at least one individual NCF at or below the cutoff. Thus the gate selected paths, not a guarantee that every waveform drawn from a selected path exceeded the threshold. This definition differs from that used for phase picking, so the numerical thresholds are not physically comparable across the two tasks.

The finite-SNR training pool contained 33,098 samples. The first and second SNR tertiles were 3.04 and 6.77 dB. The three training conditions were full-distribution matched, SNR$>$3.04 dB matched, and SNR$>$6.77 dB matched, each containing 11,033 samples. The full and SNR$>$3.04 dB pools were resampled for each seed; the SNR$>$6.77 dB pool contained exactly 11,033 samples and was reused while model initialization and NCF draws varied. Training and matched-subset sampling used seeds 20260609, 20260610, and 20260611.

All models were evaluated against the same 8,292 unfiltered test samples and the same 2,734 test samples above 6.77 dB, with identical labels, validity masks, and period grids within each test scope. Models were trained for 5 epochs with batch size 256 and an AdamW learning rate of $2\times10^{-4}$. The prespecified fifth-epoch checkpoints were used for comparison; no validation-based checkpoint selection was performed. Archived logs show that training loss fell from epoch 4 to epoch 5 for every seed in both the full-distribution and strict-SNR conditions. Interim validation metrics were not recorded in this original run. A separate, independently validated duration sensitivity is detailed in Text S12; it does not convert the original fifth-epoch checkpoints into converged models.

\subsection*{Text S3. Continuous association parameters.}
The continuous association experiment used picker outputs from 6 July 2019, the 2019 Ridgecrest mainshock day, and 13 November 2021, a 2021 comparison day. For every automatic pick, the tested streams measured $r=\sigma_\mathrm{post}/\sigma_\mathrm{pre}$ from adjacent 2 s windows. The Vertical-S definition used vertical-component standard deviations for both phases. The Horizontal-S definition retained vertical P measurement and used the horizontal-vector amplitude $[(\sigma_E^2+\sigma_N^2)/2]^{1/2}$ for S. This component comparison tests definition dependence and is not proposed as a new SNR estimator. Because $r$ is an amplitude ratio, the reported window-amplitude proxy is $20\log_{10}r$. The immediately preceding S window can include P-wave coda and is therefore not necessarily a stationary-noise estimate; a reference-timed pre-P diagnostic is reported only as a nondeployable implementation sensitivity in Text S9.

PnSn v3 was loaded from the released 5120-sample TorchScript archive. It contains a convolutional encoder, two bidirectional long short-term memory blocks, and decoder outputs for background, Pg, Sg, Pn, and Sn. Three-component windows were demeaned, divided by their maximum absolute amplitude, and evaluated at 100 Hz in 51.2 s windows with a 40.96 s stride. The internal candidate extractor used a minimum phase probability of 0.1 and a 300-sample (3 s) within-phase minimum separation. The retention comparisons added no second non-maximum-suppression stage or association-stage probability cutoff.

Reference labels and catalog origins were read from the SeismicX-Cont source archive rather than generated by this PnSn v3 inference run. The two selected days contain 30,560 manual S labels, 23,516 manual P labels, and 18,780 archive-supplied automatic P labels after coverage filtering. Thus the primary S-recall result contains no PnSn-generated reference S labels. The 2,340 catalog events use preferred Southern California Seismic Network (2,246 events) or Northern California Seismic Network (94 events) origins. We cannot exclude overlap between historical model-training data and all source waveforms; the relevant independence claim is that the reported reference arrivals and origins were not produced by the tested inference workflow.

The Vertical-S stream used P$\geq$10.00 dB and S$\geq$7.451 dB, whereas the Horizontal-S stream used P$\geq$10.00 dB and S$\geq$8.142 dB. In each SNR stream, the S threshold was calibrated from the complete two-day candidate set to match the 265,962 P picks retained by the fixed P threshold. The same cutoffs were then applied to both dates without day-specific adjustment. Each SNR stream therefore retained 265,962 P and 265,962 S picks.

The Probability-ranking comparator independently retained the top 265,962 candidates within P and S. Its minimum selected probabilities were 0.390409 and 0.318769, respectively. All three streams consequently contained 531,924 picks with identical coarse phase composition, while the retained times, stations, and event support were allowed to differ.

Each filtered stream was associated with REAL using 15 min chunks for computational tractability. All conditions used the same REAL control strings: \path{0.4/25/0.05/3/5} for \texttt{-R} and \path{4/2/3/2/1.0/0.1/1.0} for \texttt{-S}. Event recall was evaluated against the two-day catalog using one-to-one matching with 5 s origin-time and 30 km epicentral-distance tolerances; 2 s/10 km and 3 s/20 km matches were added as robustness checks.

The vertical-S score was stored in the source archive as $10\log_{10}r$; multiplying the coordinate and thresholds by two corrected the amplitude-dB convention without changing ranks, retained sets, or association outcomes. The main text uses the equal-P/S vertical-S workflow in Table S6. A historical common-threshold sensitivity is retained in Tables S1 and S3--S5 for audit continuity: SNR$\geq$8.50 dB retained 576,875 picks, comprising 369,896 P (64.1\%) and 206,979 S (35.9\%). Its global total-count probability control retained the top 576,875 picks, and a second control matched the vertical-S stream's P/S composition exactly. These values are not the primary manuscript budget.

\subsection*{Text S4. Paired bootstrap parameters.}
For sample-level robustness checks, the deterministic seed-20260609 checkpoints were re-evaluated and per-window phase TP/FP/FN counts and per-sample dispersion error sums were exported. Phase and dispersion confidence intervals used 10,000 paired bootstrap resamples over the shared 10,000 phase windows and 8,292 dispersion samples, respectively. For the Horizontal-S-minus-Vertical-S contrast, S-arrival recall was resampled over the 44 reference-event origin-hour blocks containing S labels; event precision, recall, and F1 were resampled over all 48 monitoring-hour blocks, assigning matched and missed events by catalog origin and catalog-unmatched detections by predicted origin. The differences and 95\% intervals were +0.324 [0.314, 0.335] for S-arrival recall, -0.007 [-0.035, 0.027] for event precision, +0.106 [0.091, 0.123] for event recall, and +0.055 [0.030, 0.079] for event F1. The separate Horizontal-S-minus-Probability-ranking event-recall difference was +0.063 [0.035, 0.085] when resampling the 46 catalog-origin-hour blocks containing events.

Under retrospective phase-specific count matching inside each 15 min window, the SNR-minus-probability event-recall differences were -0.017 [-0.035, 0.002], -0.020 [-0.040, 0.001], and -0.021 [-0.042, -0.001] at 2 s/10 km, 3 s/20 km, and 5 s/30 km matching, respectively. At 5 s/30 km, the mainshock-day difference was -0.013 [-0.033, 0.009], whereas the 67-event comparison-day difference was -0.284 [-0.397, -0.181]. These intervals are wider than event-level intervals because events sharing a time block are not treated as independent.

For the component/window diagnostic, we used 5,000 cluster-bootstrap resamples over events and stations separately. The event- and station-cluster intervals for horizontal-adjacent minus vertical-adjacent SNR were [3.341, 3.541] and [3.217, 3.686] dB. Corresponding intervals for horizontal pre-P minus horizontal adjacent were [16.746, 18.179] and [16.677, 18.255] dB. All bootstrap intervals describe stability within the analyzed dates, events, stations, or samples; none represents cross-date or cross-region uncertainty.

\subsection*{Text S5. Additional tabulated uncertainty and association summaries.}
Table S6 reports the Vertical-S equal-P/S comparison. Tables S13 and S14 report the Horizontal-S definition and its local-budget comparison, and Table S15 isolates component choice over identical candidate IDs; Table S1 preserves the common-threshold Vertical-S result and its two Probability-ranking controls. All rows within a comparison use identical REAL settings. Pick recall is a manual-plus-automatic coverage-filtered retained-label diagnostic, whereas event metrics are evaluated against the two-day catalog. Event precision and F1 are therefore catalog referenced. An incomplete catalog can classify uncataloged real events as false positives, and the size of that bias need not be identical across streams; using one catalog ensures a common reference but does not recover true operational precision. Table S2 separates seed-level variability from the fixed-model paired bootstrap intervals reported in the main text. The seed-level differences are paired by the same random seed and are descriptive because only three seeds were run.

\subsection*{Text S6. Event-level retained-geometry robustness and local threshold sensitivity.}
The main retained-geometry analysis uses all 2,340 catalog events under the Horizontal-S definition and equal-P/S budget, bins the difference in S-contributing stations (Probability ranking minus Horizontal-S), and compares paired recovery within each bin. This avoids defining the analysis population by the outcome being explained. Under this comparison, 999 events had at least three more S-contributing stations under the Horizontal-S definition; their Horizontal-S and Probability-ranking recovery fractions were 0.753 and 0.591. Forty-two events had at least three more S-contributing stations after Probability ranking; their corresponding recovery fractions were 0.571 and 0.929.

For completeness, the outcome-stratified vertical-S diagnostic was repeated with 1.0, 1.5, and 2.0 s true-positive pick-matching tolerances. For the 339 catalog events recovered only from the probability-retained stream, the median probability-minus-SNR difference remained +2 matched S picks and +2 S-contributing stations across all three tolerances (Table S3). Day-stratified checks gave the same direction: probability-only events on 6 July 2019 had median differences of +2.5 S picks and +2 S stations, and those on 13 November 2021 had differences of +2 and +2.

We also repeated the retained-geometry calculation over a local SNR-threshold band. For each SNR threshold, the comparator stream was the top-$N$ phase-probability stream with exactly the same retained-pick budget. This check varies the retained-pick budget and probability cutoff locally, but it does not rerun REAL association or redefine recovery classes. Across SNR thresholds of 8.00, 8.50, and 9.00 dB, the median probability-minus-SNR difference for probability-only recovered events remained +2 matched S picks and +2 S-contributing stations (Table S4).

As a stricter check, we also reran the complete 15 min REAL association workflow at the two adjacent thresholds, SNR$\geq$8.00 dB and SNR$\geq$9.00 dB, again with an exactly count-matched top-probability stream and identical REAL and event-matching parameters. The baseline 8.50 dB association is included for comparison. Across all three thresholds, the SNR stream recovered fewer catalog events than the count-matched probability stream (Table S5). Event-recovery ordering was also evaluated separately by day and under 2 s/10 km, 3 s/20 km, and 5 s/30 km event matching (Table S9). This is a local association check around the manuscript operating point, not a broad operational threshold sweep.

\subsection*{Text S7. Vertical-only continuous-filter sensitivity.}
We repeated the two-day continuous filtering and REAL-association workflow with the vertical component used for both phases, phase-specific SNR thresholds, and exact phase-specific comparator budgets. P thresholds were fixed at 10 and 20 dB, and S thresholds were chosen from the same continuous pick stream so that each SNR stream retained equal P- and S-pick counts. The resulting thresholds were S$\geq$7.451 dB for the moderate condition and S$\geq$12.024 dB for the strict condition. Within P and S separately, picks were ranked by phase probability and retained until their counts exactly matched the paired SNR stream. The moderate pair therefore contained 265,962 P and 265,962 S picks in each stream; the strict pair contained 50,691 P and 50,691 S picks. The minimum selected confidence values were P/S=0.390409/0.318769 and 0.716138/0.559599.

All four streams were divided into the same 15 min windows and associated with the same REAL control strings as the common-threshold baseline: \path{0.4/25/0.05/3/5} for \texttt{-R} and \path{4/2/3/2/1.0/0.1/1.0} for \texttt{-S}. Event recovery used the same one-to-one 5 s origin-time and 30 km epicentral-distance matching. This design fixes total pick count, coarse P/S composition, chunking, associator settings, and event matching; it varies the within-phase retention rule. At the moderate budget, SNR filtering traded lower event recall for higher catalog-referenced precision (0.536/0.651) relative to probability ranking (0.580/0.509). At the strict budget, the corresponding recall/precision values were 0.106/0.926 and 0.274/0.669 (Table S6).

We also selected probability-ranked picks separately by phase inside every 15 min chunk, exactly matching each common-threshold vertical-SNR chunk's P and S counts. This retrospective sensitivity prevents the global rank from reallocating pick budget among association windows; it is not proposed as an online threshold. The window-stratified comparator retained the same 576,875 picks, increased P/S recall from 0.771/0.357 to 0.778/0.564, and increased catalog-event recall from 0.556 to 0.662 while catalog-referenced precision changed from 0.472 to 0.485 (Table S10). The SNR-minus-probability event-recall difference was -0.106 with an origin-hour block-bootstrap interval of [-0.133, -0.081].

\subsection*{Text S8. Label-independent case-event repicking.}
We independently repicked a short waveform window for moderate probability-only event ci37227796 (origin time 2019-07-06 05:47:08.6 UTC; 35.7132$^\circ$N, 117.5407$^\circ$W; depth 1.66 km; M2.71). This event was one of two probability-only events tied for the largest matched probability-minus-SNR S-support difference (+10). The record sections in Figure S1 are therefore illustrative and are not an independent statistical test; Figure 8 of the main text instead analyzes all catalog events. We queried the SeismicX-Cont coverage index for every HH, BH, HN, or EH station-family record within 500 km that overlapped -20 to +170 s relative to the origin. A record required a complete three-component family and at least 95\% temporal coverage on each component. Of 1,156 candidate records, 1,045 records at 478 station locations met these criteria; 111 were excluded only for incomplete component coverage.

The qualified windows were resampled to 100 Hz and passed through the PnSn v3 5120-sample TorchScript picker used by the continuous workflow. The picker produced 2,797 raw picks between 0 and 160 s after the origin. We then applied the vertical-S filters without event-specific recalibration: P$\geq$10/S$\geq$7.450674 dB for the SNR stream and phase probability P$\geq$0.390409/S$\geq$0.318769 for the probability stream. These probability cutoffs exactly matched the vertical-S stream's 265,962 P and 265,962 S picks over the two complete continuous days; they were not recomputed within the selected event window. SNR was $20\log_{10}(\sigma_\mathrm{post}/\sigma_\mathrm{pre})$ from 2 s vertical-component windows before and after each automatic pick.

For visualization only, normalized vertical components were bandpass filtered at 1--10 Hz. Straight-ray curves used the REAL velocities of 6.2 km/s for P and 3.5 km/s for S. Filled symbols in Figure S1c,d fall within $\pm$5 s of the corresponding curve; this window identifies event-consistent automatic candidates and is not a reference label or an association criterion. Probability filtering retained 736 picks in the displayed window, including 99 P and 126 S picks near the curves and near-curve S picks at 61 station locations. SNR filtering retained 381 picks, including 44 P and 48 S picks near the curves and near-curve S picks at 30 station locations. Thus panels c,d use all coverage-qualified waveforms and no reference picks. The two-day phase matching was deliberately not repeated within this event window, so Figure S1 shows the local consequence of fixed global gates rather than a locally equal-budget comparison.

\begin{figure}[!htbp]
\centering
\includegraphics[width=\textwidth]{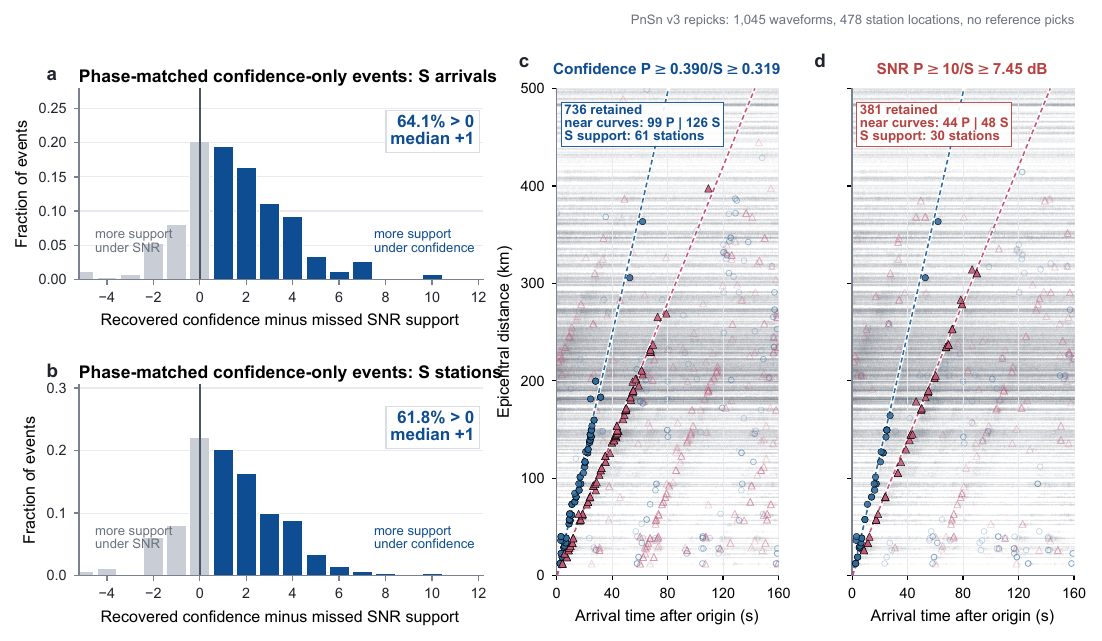}
\caption{Illustrative, outcome-selected event ci37227796 under the vertical-S implementation. Panels a and b summarize the moderate probability-only event class, and panels c and d show label-independent PnSn v3 repicking of all coverage-qualified waveforms within 500 km. Because the case was selected for a large S-support difference, it is not used as population-level mechanism evidence; Figure 8 in the main text provides that test.}
\label{si:fig:si_case_event}
\end{figure}

\subsection*{Text S9. S-arrival component and noise-window sensitivity.}
We tested whether the continuous workflow's use of a vertical component and an immediately preceding noise window structurally lowered S-arrival SNR. The diagnostic used the held-out CREDIT-X1local test split, retaining manual Sg or Sn arrivals with a preceding P arrival and complete east, north, and vertical waveforms. Of 8,396 eligible records, 8,374 S arrivals had all required windows. For each S arrival we computed $20\log_{10}(\sigma_\mathrm{signal}/\sigma_\mathrm{noise})$ with 2 s windows under three definitions: vertical signal and adjacent pre-S noise, horizontal-vector signal and adjacent pre-S noise, and horizontal-vector signal with noise measured in a 2 s window ending 2 s before the nearest preceding P arrival. Horizontal standard deviation was $[(\sigma_E^2+\sigma_N^2)/2]^{1/2}$. The pre-P definition uses reference P times and is a sensitivity diagnostic, not a deployable filter used in the association experiment.

The median horizontal-adjacent minus vertical-adjacent difference was +3.449 dB, with event- and station-cluster 95\% bootstrap intervals of [3.341, 3.541] and [3.217, 3.686] dB and 88.1\% positive differences. Moving the horizontal noise window from immediately before S to before P added a median +17.424 dB, with event- and station-cluster intervals of [16.746, 18.179] and [16.677, 18.255] dB and 98.1\% positive differences. At a common 8.50 dB threshold, 35.4\%, 70.3\%, and 97.0\% of arrivals passed the three definitions, respectively (Table S8). Because numerical scales differ, we also fixed the retained count to the 2,965 arrivals passing the vertical-adjacent gate: selected-set Jaccard overlap was 0.511 for horizontal-adjacent and 0.300 for horizontal-pre-P ranking. Thus both the SNR scale and the identity of retained S arrivals depend strongly on component and noise-window choices.

\subsection*{Text S10. Observation-geometry shifts under training curation.}
The manual phase-picking record distributions were first compared in the threshold-passing candidate pools so that censoring imposed by the gates remained visible. Median source--station distance decreased from 113.8 km in the full manual pool to 101.8 km at P/S=14/7.27 dB and 94.9 km at P/S=16/16.60 dB. Figure 2a displays 0--450 km without renormalizing the visible bins: 239 of 61,429 full-pool records (0.39\%), 97 of 45,880 lower-gate records (0.21\%), and 81 of 39,718 strict-gate records (0.20\%) lie beyond the plotted range.

The seed-fixed final training manifests show that the distance shift remained after equal-size sampling. Across the three full-distribution phase subsets, median distance ranged from 110.2 to 110.4 km, compared with 101.3--101.6 km for the lower-gate subsets and 94.9 km for every strict-gate subset. The corresponding dispersion medians were 422.9--424.8 km for full-distribution subsets, 350.5--351.7 km above 3.04 dB, and 305.6 km above 6.77 dB. Figure 3a now shows the complete finite-SNR path distribution to 1,300 km; in the final training sets, the fractions beyond 550 km were 31.1--31.8\%, 16.7--16.8\%, and 10.5\%, respectively. These statistics characterize the actual training inputs. They are not used to explain the measured F1 or dispersion-error differences.

\subsection*{Text S11. Component and eligibility sensitivity.}
For the deployable component sensitivity, P picks were held fixed to the moderate vertical-window stream. All 1,297,032 candidate S picks were rescored from the complete three-component continuous waveforms using the horizontal-vector standard deviation, $[(\sigma_E^2+\sigma_N^2)/2]^{1/2}$, in the same adjacent 2 s pre- and post-pick windows. Of these candidates, 1,293,449 had complete valid horizontal windows. As an implementation check, recomputing the original vertical score for 10,000 candidates differed from the stored score by a median of $4.3\times10^{-8}$ dB and a 95th-percentile absolute difference of $4.5\times10^{-5}$ dB.

The candidate archive used for Probability ranking contained 1,697,829 S picks. It included stations with only a vertical channel, represented by repeated vertical inputs. Among the 265,962 globally selected probability-ranked S picks, 208,192 (78.3\%) had distinct east, north, and vertical channel metadata; the remaining 57,770 could not enter a valid horizontal-component SNR ranking. In the 15 min phase-matched probability sensitivity, 222,817 (83.8\%) of 265,962 selected S picks had distinct three-component metadata. The Horizontal-S and Probability-ranking definitions therefore evaluate operationally available workflows at identical output budgets, not two scores ranked over identical eligible candidate IDs. We used a separate common-eligibility component sensitivity to isolate Vertical-S from Horizontal-S measurement.

We first performed a direct component intervention independently inside each of the 192 15 min association windows. In every window, the horizontal stream retained exactly the vertical stream's number of S picks; P picks were identical. Both streams consequently contained 265,962 P and 265,962 S picks. No reference arrival or event time entered the score or selection. We reran every window through REAL with the same parameters and evaluated the same two-day catalog. Horizontal measurement increased S-arrival recall from 0.394 to 0.662 and catalog-event recall from 0.536 to 0.645, while P recall remained 0.691 and catalog-referenced precision changed only from 0.651 to 0.650 (Table S11).

For the stricter sensitivity, both rankings were restricted to the same 1,292,360 S-candidate IDs with valid stored vertical scores and valid east, north, and vertical waveform windows. The 265,962 P records were byte-identical across streams, and both rankings retained the same S count in every 15 min window. Their selected S sets shared 146,296 IDs (Jaccard overlap 0.379), so component choice changed ranking within a fixed eligibility pool rather than by admitting different candidates. Horizontal measurement increased S-arrival recall from 0.473 to 0.662, catalog-event recall from 0.571 to 0.645, and event F1 from 0.618 to 0.648, while catalog-referenced precision decreased from 0.674 to 0.650 (Table S15). Thus the component-dependent recall ordering did not arise from different candidate eligibility.

Across all 2,340 catalog events, the Horizontal-S definition retained a median of two additional S-contributing stations; differences were positive, zero, and negative for 77.7\%, 18.7\%, and 3.5\% of events. It recovered 305 events missed by the Vertical-S definition, compared with 51 recoveries unique to Vertical-S. The Horizontal-S-minus-Vertical-S event-recall difference was +0.109 with a 95\% origin-hour block-bootstrap interval of [0.091, 0.132]. The ordering persisted under each event-matching tolerance and on each date; the interval for the 67-event comparison day included zero (Table S12). This is the fixed-threshold component comparison reported in the main text.

For the fixed-threshold Horizontal-S definition, the P threshold remained 10.00 dB and a single S threshold of 8.142 dB was calibrated over the combined candidate stream to retain the same number of S picks as P picks. Both this stream and the within-phase Probability-ranking definition retained 265,962 P and 265,962 S picks before identical REAL association. Table S13 reports event-matching and day sensitivity. Table S14 reports the additional retrospective comparison in which probability-ranked picks were selected separately by phase within each 15 min window to match the Horizontal-S counts in that window.

\subsection*{Text S12. Extended-training and validation sensitivity.}
For phase picking, we preserved the original three seed-specific full and strict training subsets, record-level OR rule, manual targets, initialization, batch order, optimizer, and learning rate. Fine-tuning continued from 2,000 to 4,000 steps and scratch training from 10,000 to 20,000 steps. The 12 model snapshots at the original endpoints (three seeds, two initializations, two subsets) were each compared with the archived checkpoint tensor by tensor; all 120 tensors per snapshot were exactly equal. Both original and extended endpoints were then scored on the same deterministic 10,000-window unfiltered and strict-SNR manual-label tests for each seed, with no checkpoint chosen from test performance. Each evaluation was repeated twice with identical metrics. Table S17 reports seed-paired strict-minus-full pooled P/S F1, with P and S contingency counts combined before F1 calculation. At 20,000 scratch steps on the unfiltered test, the mean strict-minus-full P-wave F1 was +0.00543 and S-wave F1 was -0.00589; their opposing directions help explain the small pooled difference. These longer runs test schedule sensitivity but do not establish convergence because no independent phase-picking validation trajectory was used to select or stop training.

We therefore ran a separate validation-guided phase sensitivity rather than choosing a longer endpoint from the test results. The original 61,429-record training split was separated by event key into 55,247 fit records (4,191 events) and 6,182 validation records (466 events); neither shared events with the original 16,181-record test. The strict P/S thresholds were recalibrated on fit records alone to 16.0/16.586 dB, balancing threshold-passing phase labels there. Under the same record-level OR rule, each final full or strict training set contained 35,661 records, comprising 10,372 P-only, 24,304 P+S, and 985 S-only records. For every seed, full and strict models used equal batch size, learning rate, architecture, and number of update steps. A fixed 3,000-window sample of unfiltered validation records was scored at 2,000-step intervals through 8,000 fine-tuning steps and at 4,000-step intervals from 8,000 through 28,000 scratch steps. The prespecified stopping check required neither condition's three-seed mean pooled P/S validation F1 to improve by at least 0.002 for two successive checks. One checkpoint was then chosen for both conditions and all three seeds within each initialization mode: the earliest within 0.002 of the best mean of the full- and strict-condition validation scores. The original test was accessed only after this common step was fixed; holdout scores used the same deterministic 8,000 windows per test scope for all models. Fine-tuning used verified, identical archived trajectories and validation metrics from the fit--validation split, avoiding duplicate training; scratch models were trained anew on the same manifests.

Fine-tuning reached the 8,000-step compute cap while mean validation F1 was still increasing in both conditions. Scratch mean validation F1 peaked at 24,000 steps (full 0.7170; strict 0.7054) and fell at 28,000 (full 0.7148; strict 0.7024). The latter was only one stale check, so neither mode met the two-check stopping rule. The common validation-selected steps were 8,000 for fine-tuning and 24,000 for scratch. They are budget-limited, validation-selected comparisons, not convergence estimates. Table S19 reports test results as paired strict-minus-full differences; these values cannot be directly pooled with Table S17 because the fit split, budget, thresholds, and deterministic test-window sample differ.

For dispersion, we first reevaluated the original fifth-epoch seed-20260609 checkpoints on CPU without retraining. Their unfiltered-test MAE values, 0.046204484 km s$^{-1}$ for full and 0.050973617 km s$^{-1}$ for strict, exactly reproduced the archived summaries. The duration sensitivity was then trained independently: 3,310 of the original 33,098 finite-SNR training paths were withheld as a station-pair-disjoint validation split, leaving 29,788 fit paths. The strict cutoff was recalculated only within that fit pool (6.777488 dB), and both conditions used 9,929 training paths. The original 8,292-path unfiltered test remained untouched until endpoint scoring; 2,730 of its paths passed the recalibrated strict cutoff, compared with 2,734 under the original cutoff. Three matched seeds ran for 15 epochs with batch size 256, AdamW learning rate $2\times10^{-4}$, and a cosine schedule over 15 epochs. Epochs 5, 10, and 15 were specified before test evaluation. A separate checkpoint for each run minimized MAE on a fixed 512-path unfiltered validation subset; the chosen epochs were 12, 13, and 13 for full and 13, 12, and 11 for strict. Deterministic test inputs were scored on CPU, and no training batch required a nonfinite-value retry. Table S18 reports paired epoch-specific MAE differences. The changed fit split, matched budget, strict-test membership, and learning-rate schedule make this a new experiment rather than a direct continuation of the original five-epoch models. The validation-selected results also favored strict training on the unfiltered test in all three pairs, but the third pair differed by only 0.000047 km s$^{-1}$; neither this result nor a near-flat validation curve proves asymptotic convergence.

A retrospective audit of the six saved learning curves checked the common, prespecified epoch-15 comparison without retraining or selecting a checkpoint by test performance. The mean unfiltered-validation MAE across all six models reached 0.040403 km s$^{-1}$ at epoch 13 and was 0.040438 km s$^{-1}$ at epoch 15, a difference of 0.000035 km s$^{-1}$. At epoch 15, the three-seed mean was 0.040899 for full training and 0.039977 km s$^{-1}$ for strict-SNR training. Thus the reported common epoch is close to the minimum observed under this 15-epoch schedule. This post hoc audit does not establish a prospective early-stopping rule or asymptotic convergence; Table S18 remains a common fixed-epoch sensitivity, not a test-selected optimum.

Both duration sensitivities show that the predictive ordering changes with the training protocol. Text S13 adds a phase-picking comparison at matched distance-bin counts and phase composition to test whether strict screening adds value under those controls; no analogous dispersion control was trained. Neither comparison identifies a mechanism for the performance differences.

\subsection*{Text S13. Joint distance and pre-gate SNR-distribution control for phase picking.}
We used the event-disjoint fit, validation, and holdout splits from Text S12, without changing the validation-selected 8,000-step fine-tuning endpoint. In the 55,247-record fit pool, the recalibrated strict gate retained 35,661 records using P=16.0 and S=16.586 dB and the same P-or-S rule. For each fit record, we defined a record-level SNR margin as the largest available manual P or S SNR minus its phase-specific threshold. Within each P-only, P+S, or S-only record stratum, a control subset was sampled from the pre-gate fit pool to match the strict subset's counts in 25 source--station distance quantile bins and the pre-gate pool's counts in 10 SNR-margin quantile bins. Integer allocations were solved jointly across the two bin dimensions; records within each cell were sampled separately for each of the three existing seeds. Both subsets contained 10,372 P-only, 24,304 P+S, and 985 S-only records. No validation or holdout record entered matching, and the holdout was scored once at the previously selected common step.

The control's median distance was 94.5--94.6 km across seeds versus 94.35 km for strict selection; distance Kolmogorov--Smirnov statistics were 0.00165--0.00182. Because the control retained the broad pool's SNR-margin bins rather than imposing the hard gate, 72.7\% of its records also belonged to the strict subset. On the unfiltered holdout, strict-minus-control pooled P/S F1 was +0.00356, -0.00123, and -0.00134 by seed (mean +0.00033), with no consistent advantage. On the strict-SNR holdout the differences were +0.00575, +0.00323, and +0.00485 (mean +0.00461), favoring strict training on its selected domain (Table S20). These are descriptive three-seed contrasts, not equivalence tests. The control matches binned distance and phase composition while preserving coarse pre-gate SNR-margin frequencies; it does not equate continuous P and S SNR values, station or event distributions, or other waveform properties. Its substantial record overlap with the strict subset also limits sensitivity to differences among the nonshared records. This comparison therefore tests the added broad-test value of the hard gate against one physically matched alternative, not a complete causal decomposition of the original full-versus-strict contrast.

\subsection*{Text S14. Full-available dispersion training at a fixed update budget.}
We added a full-available condition to the independent-validation dispersion experiment in Text S12. It used all 29,788 finite-SNR fit paths, with the same 3,310 validation paths and 8,292 original test paths excluded from training. The strict subset contained 9,929 paths above 6.777488 dB, and its test subset contained 2,730 paths. The two archived 9,929-path conditions (matched full and strict) and the new full-available models shared seeds 20260609, 20260610, and 20260611, the DispNet v2.3 architecture, optimizer, loss weights, and original data-loading rules. The reference dispersion curve was estimated from each condition's training paths, as in the archived workflow.

To retain the existing compute budget, training was organized into 15 blocks of 39 updates. Every block used 38 batches of 256 paths and one of 201, for exactly 585 updates and 148,935 path presentations per model. The learning rate followed the archived 15-block cosine schedule from $2\times10^{-4}$ toward $10^{-6}$, changing only between blocks. In the new condition, all fit paths were shuffled in continuous cycles that were not reset at block boundaries. Every path was presented four or five times, whereas the smaller conditions presented each path 15 times. The same prespecified 512-path unfiltered validation subset was scored after each block. The primary comparison used block 15, locked before the new training; saving the minimum-validation checkpoint did not change that endpoint. All three new runs finished before holdout scoring. All nine frozen models were scored with identical deterministic first-NCF inputs and the existing batchwise metrics. Accelerated MPS inference reproduced all archived CPU metrics within $1.9\times10^{-9}$. Each new model also passed a first-batch CPU comparison within $10^{-6}$, and repeated MPS scoring returned identical metrics. This inference-only acceleration changed neither training nor the selected checkpoint.

Full-available training reduced mean unfiltered-test MAE by 0.001251 km s$^{-1}$ (3.0\%) relative to matched full training, with paired reductions of 0.001790, 0.001439, and 0.000525 for seeds 20260609--20260611. The corresponding strict-test reduction averaged 0.001140 km s$^{-1}$ (2.7\%). Strict-minus-full-available unfiltered-test MAE was -0.000212, +0.000053, and +0.000425 by seed (mean +0.000088), giving mixed ordering and near-equal means (Table S21). On the strict-SNR test, strict training had lower MAE in every seed, by an average 0.004121 km s$^{-1}$ (9.9\%). Mean unfiltered-test RMSE was 0.058524, 0.056135, and 0.056191 km s$^{-1}$ for matched full, strict, and full available, respectively; strict-test RMSE was 0.058458, 0.051321, and 0.056370 km s$^{-1}$. The small unfiltered strict-versus-full-available ordering therefore also depends on which error measure is used. These three-seed descriptions do not establish statistical equivalence or cross-archive effects.

This design distinguishes unique training paths from total presentations. Full available versus matched full tests the value of additional eligible paths at fixed compute; strict versus full available retains both the composition change and the data loss caused by screening. It does not compare separate, independently optimized training schedules, and the larger set necessarily receives fewer repetitions per path. Whole-path retention still selects the path-level median SNR described in Text S2, rather than applying a threshold to every NCF drawn during training.

\begin{figure}[!htbp]
\centering
\includegraphics[width=\textwidth]{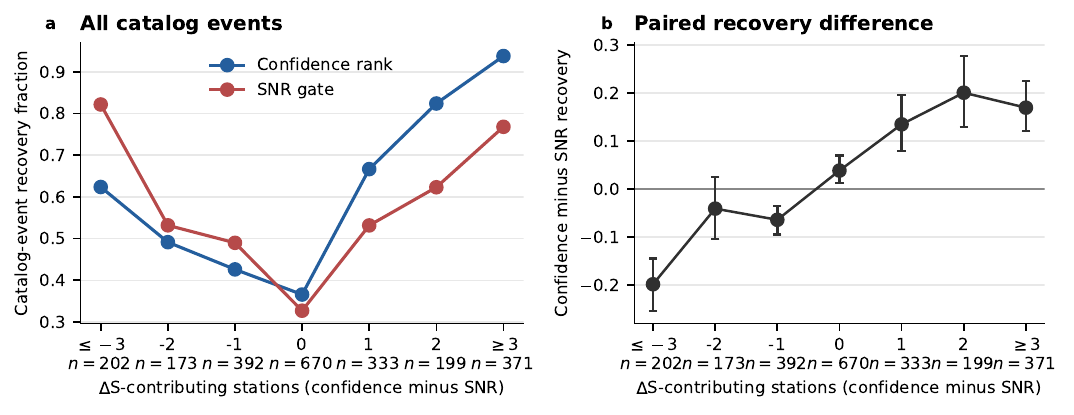}
\caption{All-event retained-support relation for the moderate equal-P/S Vertical-S definition. Events are binned by the Probability-ranking-minus-Vertical-S difference in retained S-contributing stations. (a) Recovery fraction under each rule. (b) Paired Probability-ranking-minus-Vertical-S recovery difference with 95\% origin-hour block-bootstrap intervals. The Horizontal-S definition is shown in Figure 8 of the main text.}
\label{si:fig:si_vertical_support}
\end{figure}

\begin{table}
\caption{Historical vertical-only common-threshold sensitivity on two continuous-data days. Both probability controls match the SNR stream's total pick count; the within-phase control additionally matches its P/S counts. These 576,875-pick streams are not the primary equal-P/S manuscript budget. Pick recall is manual-plus-automatic coverage-filtered recall. Event predicted count, catalog-referenced precision, recall, and F1 are measured against 2,340 catalog events after identical REAL association and 5 s / 30 km event matching.}
\label{si:tab:si_continuous_assoc}
\centering
\scriptsize
\setlength{\tabcolsep}{2pt}
\begin{adjustbox}{max width=\linewidth}
\begin{tabular}{p{0.19\textwidth}rrrrrrrrr}
\hline
Filter & Picks & P rec. & S rec. & Pick rec. & Event pred. & Event TP & Event prec. & Event rec. & Event F1 \\
\hline
SNR filter & 576,875 & 0.771 & 0.357 & 0.601 & 2,755 & 1,301 & 0.472 & 0.556 & 0.511 \\
Global rank & 576,875 & 0.726 & 0.619 & 0.682 & 3,182 & 1,561 & 0.491 & 0.667 & 0.565 \\
Within-phase rank & 576,875 & 0.764 & 0.540 & 0.672 & 3,250 & 1,553 & 0.478 & 0.664 & 0.556 \\
\hline
\end{tabular}
\end{adjustbox}
\end{table}

\begin{table}
\caption{Seed-level paired differences on the unfiltered tests relative to full-distribution training. Phase-picking differences use pooled P/S F1 and are reported separately for 2,000-step fine-tuning and 10,000-step scratch initialization. Negative F1 differences and positive MAE differences indicate lower pooled F1 and larger dispersion error, respectively. With only three seeds, these summaries describe run-to-run stability rather than statistical equivalence or a population effect.}
\label{si:tab:si_seed_differences}
\centering
\scriptsize
\setlength{\tabcolsep}{3pt}
\begin{adjustbox}{max width=\linewidth}
\begin{tabular}{p{0.16\textwidth}p{0.42\textwidth}p{0.18\textwidth}p{0.13\textwidth}}
\hline
Task & Comparison & Mean paired difference & Sample SD \\
\hline
Phase picking & Fine-tune P14/S7.27 dB pooled F1 minus full & -0.00516 & 0.00412 \\
Phase picking & Fine-tune P16/S16.60 dB pooled F1 minus full & -0.00804 & 0.00419 \\
Phase picking & Scratch P14/S7.27 dB pooled F1 minus full & -0.00534 & 0.03298 \\
Phase picking & Scratch P16/S16.60 dB pooled F1 minus full & -0.02094 & 0.03735 \\
Dispersion & SNR$>$3.04 dB MAE minus full & 0.00167 km/s & 0.00158 km/s \\
Dispersion & SNR$>$6.77 dB MAE minus full & 0.00476 km/s & 0.00117 km/s \\
\hline
\end{tabular}
\end{adjustbox}
\end{table}

\begin{table}
\caption{Vertical-only SNR retained-geometry robustness for the 339 catalog events recovered only from the probability-retained stream. Differences are probability minus SNR for the same reference events. Positive fractions give the fraction of events with more probability-retained support than SNR-retained support. Recovery classes are fixed from the association comparison.}
\label{si:tab:si_geometry_tolerance}
\centering
\scriptsize
\setlength{\tabcolsep}{2pt}
\begin{adjustbox}{max width=\linewidth}
\begin{tabular}{lrrrr}
\hline
Pick tolerance & Median $\Delta$S picks & Positive $\Delta$S picks & Median $\Delta$S stations & Positive $\Delta$S stations \\
\hline
1.0 s & 2.0 & 82.6\% & 2.0 & 82.0\% \\
1.5 s & 2.0 & 82.6\% & 2.0 & 82.0\% \\
2.0 s & 2.0 & 83.2\% & 2.0 & 82.6\% \\
\hline
\end{tabular}
\end{adjustbox}
\end{table}

\begin{table}
\caption{Vertical-only SNR local-threshold retained-geometry sensitivity for the probability-only recovery class. Each probability comparator is count matched to the SNR stream at the same retained-pick budget. This is a retained-geometry sensitivity check, not a full re-association sweep.}
\label{si:tab:si_geometry_threshold}
\centering
\scriptsize
\setlength{\tabcolsep}{2pt}
\begin{adjustbox}{max width=\linewidth}
\begin{tabular}{lrrrrrr}
\hline
SNR cut & Picks & Conf. cutoff & Med. $\Delta$S picks & Pos. $\Delta$S picks & Med. $\Delta$S stations & Pos. $\Delta$S stations \\
\hline
8.00 dB & 646,981 & 0.365879 & 2.0 & 81.7\% & 2.0 & 81.4\% \\
8.50 dB & 576,875 & 0.390174 & 2.0 & 82.6\% & 2.0 & 82.0\% \\
9.00 dB & 514,635 & 0.414658 & 2.0 & 82.9\% & 2.0 & 82.3\% \\
\hline
\end{tabular}
\end{adjustbox}
\end{table}

\begin{table}
\caption{Vertical-only SNR local-threshold association check. For each SNR threshold, the probability comparator keeps the same number of picks by global phase-probability top-$N$ ranking. All rows use the same 15 min REAL chunking, REAL parameters, and 5 s / 30 km event-matching rule.}
\label{si:tab:si_association_threshold}
\centering
\scriptsize
\setlength{\tabcolsep}{3pt}
\begin{adjustbox}{max width=\linewidth}
\begin{tabular}{lrrrrrrr}
\hline
SNR cut & Picks & Conf. cutoff & SNR TP & Conf. TP & SNR rec. & Conf. rec. & $\Delta$TP \\
\hline
8.00 dB & 646,981 & 0.365879 & 1,382 & 1,571 & 0.591 & 0.671 & -189 \\
8.50 dB & 576,875 & 0.390174 & 1,301 & 1,561 & 0.556 & 0.667 & -260 \\
9.00 dB & 514,635 & 0.414658 & 1,211 & 1,542 & 0.518 & 0.659 & -331 \\
\hline
\end{tabular}
\end{adjustbox}
\end{table}

\begin{table}
\caption{Vertical-only SNR, phase-matched continuous-filter sensitivity on the same two continuous-data days. P thresholds were fixed at 10 and 20 dB; S thresholds were selected to retain equal P- and S-pick counts in the SNR stream. Within P and S separately, probability comparators retain the highest-scoring picks until their phase counts exactly match the SNR stream. Pick recall is manual-plus-automatic coverage-filtered recall, and catalog-referenced event metrics are measured against 2,340 catalog events after identical 15 min REAL association and 5 s / 30 km event matching.}
\label{si:tab:si_phase_balanced_continuous}
\centering
\scriptsize
\setlength{\tabcolsep}{2pt}
\begin{adjustbox}{max width=\linewidth}
\begin{tabular}{p{0.09\textwidth}p{0.27\textwidth}rrrrrrrr}
\hline
Budget & Retention rule & Picks & P rec. & S rec. & Pick rec. & Event TP & Event prec. & Event rec. & Event F1 \\
\hline
Moderate & SNR P$\geq$10/S$\geq$7.45 dB & 531,924 & 0.691 & 0.394 & 0.569 & 1,255 & 0.651 & 0.536 & 0.588 \\
Moderate & Probability P/S$\geq$0.390/0.319 & 531,924 & 0.615 & 0.441 & 0.543 & 1,357 & 0.509 & 0.580 & 0.542 \\
\hline
Strict & SNR P$\geq$20/S$\geq$12.02 dB & 101,382 & 0.280 & 0.096 & 0.204 & 249 & 0.926 & 0.106 & 0.191 \\
Strict & Probability P/S$\geq$0.716/0.560 & 101,382 & 0.288 & 0.133 & 0.224 & 640 & 0.669 & 0.274 & 0.388 \\
\hline
\end{tabular}
\end{adjustbox}
\end{table}

\begin{table}
\caption{Cross-evaluation using only manual phase labels. Values are pooled P/S F1 means (sample SD) across three seeds at picker threshold 0.1; P and S contingency counts are summed before F1 is calculated. Training-calibrated test gates were P$\geq$14/S$\geq$7.27 dB and P$\geq$16/S$\geq$16.60 dB and retained complete records when either phase passed. Each row compares full-distribution and paired gate training on the same unfiltered and gate-filtered test windows.}
\label{si:tab:si_snr_filtered_test_precision}
\centering
\scriptsize
\setlength{\tabcolsep}{3pt}
\begin{adjustbox}{max width=\linewidth}
\begin{tabular}{llrrrr}
\hline
Initialization & Gate & \multicolumn{2}{c}{Full test F1} & \multicolumn{2}{c}{Gate test F1}\\
 & & Full train & Gate train & Full train & Gate train\\
\hline
Transfer & P14/S7.27 & 0.739 (0.002) & 0.733 (0.005) & 0.771 (0.003) & 0.770 (0.002)\\
Transfer & P16/S16.60 & 0.739 (0.002) & 0.731 (0.006) & 0.784 (0.002) & 0.780 (0.003)\\
Scratch & P14/S7.27 & 0.687 (0.021) & 0.681 (0.019) & 0.719 (0.016) & 0.719 (0.019)\\
Scratch & P16/S16.60 & 0.687 (0.021) & 0.666 (0.019) & 0.733 (0.019) & 0.722 (0.016)\\
\hline
\end{tabular}
\end{adjustbox}
\end{table}

\begin{table}
\caption{Component and noise-window sensitivity for 8,374 held-out manual S arrivals. Passing fractions use a common 8.50 dB threshold only to illustrate scale sensitivity. Jaccard overlap fixes every alternative retained set to the 2,965 arrivals passing the vertical-adjacent definition.}
\label{si:tab:si_snr_implementation}
\centering
\small
\setlength{\tabcolsep}{5pt}
\begin{adjustbox}{max width=\linewidth}
\begin{tabular}{lrrr}
\hline
SNR definition & Median (dB) & Pass at 8.50 dB & Fixed-budget Jaccard \\
\hline
Vertical, adjacent pre-S noise & 7.06 & 35.4\% & 1.000 \\
Horizontal, adjacent pre-S noise & 10.74 & 70.3\% & 0.511 \\
Horizontal, pre-P noise & 28.49 & 97.0\% & 0.300 \\
\hline
\end{tabular}
\end{adjustbox}
\end{table}

\begin{table}
\caption{Vertical-S definition: event-matching and day-stratified robustness. Entries are Vertical-S-minus-Probability-ranking event-recovery differences. Negative values indicate lower catalog-event recovery under the Vertical-S definition. The 2019 and 2021 dates contain 2,273 and 67 reference events, respectively.}
\label{si:tab:si_event_matching_days}
\centering
\small
\setlength{\tabcolsep}{4pt}
\begin{adjustbox}{max width=\linewidth}
\begin{tabular}{llrrr}
\hline
Budget control & Scope & 2 s/10 km & 3 s/20 km & 5 s/30 km \\
\hline
Global total count & All & -0.109 & -0.106 & -0.111 \\
Global total count & 2019-07-06 & -0.100 & -0.097 & -0.103 \\
Global total count & 2021-11-13 & -0.388 & -0.403 & -0.403 \\
\addlinespace
Original P/S counts & All & -0.102 & -0.103 & -0.108 \\
Original P/S counts & 2019-07-06 & -0.092 & -0.094 & -0.099 \\
Original P/S counts & 2021-11-13 & -0.418 & -0.418 & -0.418 \\
\addlinespace
Moderate equal P/S & All & -0.024 & -0.034 & -0.044 \\
Moderate equal P/S & 2019-07-06 & -0.014 & -0.025 & -0.034 \\
Moderate equal P/S & 2021-11-13 & -0.358 & -0.343 & -0.358 \\
\addlinespace
Strict equal P/S & All & -0.156 & -0.165 & -0.167 \\
Strict equal P/S & 2019-07-06 & -0.140 & -0.150 & -0.152 \\
Strict equal P/S & 2021-11-13 & -0.687 & -0.687 & -0.687 \\
\hline
\end{tabular}
\end{adjustbox}
\end{table}

\begin{table}
\caption{Vertical-only SNR 15 min phase-stratified count matching. The probability comparator independently matches the SNR stream's P and S counts inside every association window, preventing global top-$N$ ranking from reallocating pick budget among windows. Event metrics use identical REAL settings and 5 s/30 km catalog matching.}
\label{si:tab:si_window_phase_matching}
\centering
\scriptsize
\setlength{\tabcolsep}{2pt}
\begin{adjustbox}{max width=\linewidth}
\begin{tabular}{lrrrrrr}
\hline
Retention rule & Picks & P recall & S recall & Event precision & Event recall & Event F1 \\
\hline
Vertical-window SNR & 576,875 & 0.771 & 0.357 & 0.472 & 0.556 & 0.511 \\
Window/phase probability rank & 576,875 & 0.778 & 0.564 & 0.485 & 0.662 & 0.560 \\
\hline
\end{tabular}
\end{adjustbox}
\end{table}

\begin{table}
\caption{Direct continuous component intervention. P picks are identical, and vertical- and horizontal-S streams retain identical P and S counts inside every 15 min window. Event metrics use identical REAL settings and 5 s/30 km catalog matching.}
\label{si:tab:si_horizontal_component}
\centering
\scriptsize
\setlength{\tabcolsep}{1.5pt}
\begin{adjustbox}{max width=\linewidth}
\begin{tabular}{lrrrrrrr}
\hline
S-wave measurement & Picks & P recall & S recall & Event pred. & Event prec. & Event recall & Event F1 \\
\hline
Vertical, adjacent windows & 531,924 & 0.691 & 0.394 & 1,927 & 0.651 & 0.536 & 0.588 \\
Horizontal, adjacent windows & 531,924 & 0.691 & 0.662 & 2,322 & 0.650 & 0.645 & 0.647 \\
\hline
\end{tabular}
\end{adjustbox}
\end{table}

\begin{table}
\caption{Event-recall robustness for the direct component intervention. Entries are horizontal-minus-vertical recall differences. The final column gives the 95\% origin-hour block-bootstrap interval for the 5 s/30 km result.}
\label{si:tab:si_horizontal_component_robustness}
\centering
\small
\setlength{\tabcolsep}{5pt}
\begin{adjustbox}{max width=\linewidth}
\begin{tabular}{lrrrr}
\hline
Scope & 2 s/10 km & 3 s/20 km & 5 s/30 km & 5 s/30 km interval \\
\hline
Both dates & +0.103 & +0.103 & +0.109 & [0.091, 0.132] \\
2019-07-06 & +0.104 & +0.104 & +0.110 & [0.092, 0.133] \\
2021-11-13 & +0.090 & +0.060 & +0.075 & [-0.000, 0.177] \\
\hline
\end{tabular}
\end{adjustbox}
\end{table}

\begin{table}
\caption{Fixed-threshold Horizontal-S definition. Each Horizontal-S and Probability-ranking stream retains 265,962 P and 265,962 S picks globally. Entries report catalog-event recovery after identical REAL association; $\Delta$ is Horizontal-S minus Probability ranking. The 2019 and 2021 dates contain 2,273 and 67 reference events, respectively.}
\label{si:tab:si_primary_robustness}
\centering
\small
\setlength{\tabcolsep}{5pt}
\begin{adjustbox}{max width=\linewidth}
\begin{tabular}{llrrr}
\hline
Scope & Match tolerance & SNR recall & Probability recall & $\Delta$ recall \\
\hline
Both dates & 2 s/10 km & 0.605 & 0.532 & +0.072 \\
Both dates & 3 s/20 km & 0.626 & 0.562 & +0.064 \\
Both dates & 5 s/30 km & 0.643 & 0.580 & +0.063 \\
\addlinespace
2019-07-06 & 2 s/10 km & 0.609 & 0.524 & +0.085 \\
2019-07-06 & 3 s/20 km & 0.630 & 0.554 & +0.076 \\
2019-07-06 & 5 s/30 km & 0.647 & 0.571 & +0.075 \\
\addlinespace
2021-11-13 & 2 s/10 km & 0.463 & 0.836 & -0.373 \\
2021-11-13 & 3 s/20 km & 0.507 & 0.851 & -0.343 \\
2021-11-13 & 5 s/30 km & 0.507 & 0.866 & -0.358 \\
\hline
\end{tabular}
\end{adjustbox}
\end{table}

\begin{table}
\caption{Temporal-budget sensitivity for the Horizontal-S definition. The global rows use one fixed SNR threshold per phase and one global within-phase probability cutoff. The local rows select probability-ranked picks separately by phase in each 15 min window to match that window's Horizontal-S phase counts; the Horizontal-S stream is consequently the same count in every paired row. Event metrics use identical REAL settings and 5 s/30 km catalog matching.}
\label{si:tab:si_primary_window_budget}
\centering
\scriptsize
\setlength{\tabcolsep}{4pt}
\begin{adjustbox}{max width=\linewidth}
\begin{tabular}{llrrrrr}
\hline
Budget & Retention rule & P rec. & S rec. & Event prec. & Event rec. & F1 \\
\hline
Global & Horizontal-S definition & 0.691 & 0.718 & 0.644 & 0.643 & 0.643 \\
Global & Probability rank & 0.615 & 0.441 & 0.509 & 0.580 & 0.542 \\
\addlinespace
15 min & Horizontal-S definition & 0.691 & 0.662 & 0.650 & 0.645 & 0.647 \\
15 min & Probability rank & 0.725 & 0.628 & 0.531 & 0.666 & 0.591 \\
\hline
\end{tabular}
\end{adjustbox}
\end{table}

\begin{table}
\caption{Common-eligibility component sensitivity. Vertical and horizontal SNR rankings used the same 1,292,360 valid three-component S-candidate IDs, identical P picks, and identical S counts in every 15 min window. Each row retained 265,962 P and 265,962 S picks. Event metrics use identical REAL settings and 5 s/30 km catalog matching. Precision is catalog referenced.}
\label{si:tab:si_common_eligibility_component}
\centering
\small
\setlength{\tabcolsep}{4pt}
\begin{adjustbox}{max width=\linewidth}
\begin{tabular}{lrrrrr}
\hline
S measurement & P recall & S recall & Event precision & Event recall & Event F1 \\
\hline
Vertical & 0.691 & 0.473 & 0.674 & 0.571 & 0.618 \\
Horizontal vector & 0.691 & 0.662 & 0.650 & 0.645 & 0.648 \\
\hline
\end{tabular}
\end{adjustbox}
\end{table}

\begin{table}
\caption{Dispersion train--test cross-evaluation. Full-distribution and SNR$>$6.77 dB models used 11,033 training samples per condition and three seeds. Values are mean (sample SD) over seeds. Both model sets were evaluated on the same 8,292 unfiltered test paths and the same 2,734 test paths above 6.77 dB. Lower error is better.}
\label{si:tab:si_dispersion_cross_evaluation}
\centering
\small
\setlength{\tabcolsep}{5pt}
\begin{adjustbox}{max width=\linewidth}
\begin{tabular}{llrr}
\hline
Training distribution & Test distribution & MAE (km/s) & RMSE (km/s) \\
\hline
Full & Unfiltered & 0.0462 (0.0008) & 0.0638 (0.0023) \\
SNR$>$6.77 dB & Unfiltered & 0.0510 (0.0004) & 0.0690 (0.0008) \\
\addlinespace
Full & SNR$>$6.77 dB & 0.0464 (0.0005) & 0.0630 (0.0016) \\
SNR$>$6.77 dB & SNR$>$6.77 dB & 0.0481 (0.0007) & 0.0648 (0.0010) \\
\hline
\end{tabular}
\end{adjustbox}
\end{table}

\begin{table}
\caption{Extended phase-picking training at an unchanged matched 39,718-record budget. Entries are seed-paired strict-minus-full pooled P/S F1 on the same 10,000-window manual-label test for each seed and scope; positive values favor strict-SNR training. Original 2,000-step fine-tuning and 10,000-step scratch checkpoints were continued to 4,000 and 20,000 steps, respectively. The final column is the arithmetic mean of three paired differences, not a significance interval.}
\label{si:tab:si_extended_phase}
\centering
\scriptsize
\setlength{\tabcolsep}{4pt}
\begin{adjustbox}{max width=\linewidth}
\begin{tabular}{ll lrrrr}
\hline
Initialization & Steps & Test & Seed 09 & Seed 10 & Seed 11 & Mean \\
\hline
Fine-tune & 2,000 & Unfiltered & -0.01189 & -0.00865 & -0.00358 & -0.00804 \\
Fine-tune & 4,000 & Unfiltered & -0.00470 & -0.00111 & -0.00183 & -0.00255 \\
Fine-tune & 2,000 & Strict SNR & -0.00791 & -0.00467 & +0.00193 & -0.00355 \\
Fine-tune & 4,000 & Strict SNR & +0.00118 & +0.00331 & +0.00341 & +0.00263 \\
Scratch & 10,000 & Unfiltered & +0.01676 & -0.02166 & -0.05793 & -0.02094 \\
Scratch & 20,000 & Unfiltered & +0.00063 & +0.00317 & -0.00842 & -0.00154 \\
Scratch & 10,000 & Strict SNR & +0.02546 & -0.01432 & -0.04178 & -0.01021 \\
Scratch & 20,000 & Strict SNR & +0.01424 & +0.01421 & -0.00031 & +0.00938 \\
\hline
\end{tabular}
\end{adjustbox}
\end{table}

\FloatBarrier
\begin{table}[!ht]
\caption{Independent-validation dispersion duration sensitivity at a new matched 9,929-path training budget. Entries are strict-minus-full mean absolute error (km s$^{-1}$) at prespecified epochs; negative values favor strict-SNR training. The unfiltered test contains the original 8,292 station pairs, whereas the recalibrated strict test contains 2,730 paths. These are paired seed differences within the rerun, not direct differences from the original 11,033-path experiment.}
\label{si:tab:si_extended_dispersion}
\centering
\small
\setlength{\tabcolsep}{5pt}
\begin{adjustbox}{max width=\linewidth}
\begin{tabular}{rlrr}
\hline
Epoch & Seed & Unfiltered test & Strict-SNR test \\
\hline
5 & 20260609 & +0.000526 & -0.002965 \\
5 & 20260610 & +0.000891 & -0.000546 \\
5 & 20260611 & +0.005298 & +0.000442 \\
5 & Mean & +0.002238 & -0.001023 \\
\addlinespace
10 & 20260609 & -0.001083 & -0.005898 \\
10 & 20260610 & -0.001606 & -0.004955 \\
10 & 20260611 & +0.001003 & -0.002256 \\
10 & Mean & -0.000562 & -0.004370 \\
\addlinespace
15 & 20260609 & -0.002002 & -0.006038 \\
15 & 20260610 & -0.001386 & -0.005621 \\
15 & 20260611 & -0.000100 & -0.004124 \\
15 & Mean & -0.001163 & -0.005261 \\
\hline
\end{tabular}
\end{adjustbox}
\end{table}

\begin{table}[!ht]
\caption{Validation-guided phase-picking sensitivity at a matched 35,661-record budget. Entries are strict-minus-full pooled P/S F1 on the same deterministic manual-label holdout windows for each seed and test scope; positive values favor strict-SNR training. The common training step for all three seeds and both conditions was selected on event-disjoint validation events (8,000 fine-tuning; 24,000 scratch), not on these holdout scores. Neither mode met the prespecified two-check validation plateau rule before its compute cap. The final column is a descriptive three-seed mean, not a significance interval.}
\label{si:tab:si_phase_validation_stop}
\centering
\small
\setlength{\tabcolsep}{5pt}
\begin{adjustbox}{max width=\linewidth}
\begin{tabular}{llrrrr}
\hline
Initialization & Test & Seed 09 & Seed 10 & Seed 11 & Mean \\
\hline
Fine-tune & Unfiltered & -0.00273 & -0.00256 & -0.00269 & -0.00266 \\
Fine-tune & Strict SNR & +0.00151 & +0.00089 & +0.00126 & +0.00122 \\
Scratch & Unfiltered & -0.01296 & -0.01887 & +0.00181 & -0.01001 \\
Scratch & Strict SNR & -0.00566 & -0.01650 & +0.00416 & -0.00600 \\
\hline
\end{tabular}
\end{adjustbox}
\end{table}

\begin{table}[!ht]
\caption{Distance- and SNR-margin-matched phase-picking fine-tuning sensitivity at the common validation-selected 8,000-step checkpoint. Full and strict are the existing Text S12 runs; the joint control matches strict distance-bin counts and phase composition while retaining the eligible pre-gate pool's SNR-margin decile frequencies. All conditions contain 35,661 manual-label fit records and are evaluated on identical event-disjoint holdout windows. Values are pooled P/S F1; means summarize three seeds without implying statistical equivalence.}
\label{si:tab:si_phase_joint_control}
\centering
\small
\setlength{\tabcolsep}{5pt}
\begin{adjustbox}{max width=\linewidth}
\begin{tabular}{llrrrr}
\hline
Test & Training set & Seed 09 & Seed 10 & Seed 11 & Mean \\
\hline
Unfiltered & Full & 0.74357 & 0.74070 & 0.74236 & 0.74221 \\
Unfiltered & Strict SNR & 0.74083 & 0.73814 & 0.73967 & 0.73955 \\
Unfiltered & Joint control & 0.73727 & 0.73937 & 0.74101 & 0.73922 \\
\addlinespace
Strict SNR & Full & 0.79054 & 0.79075 & 0.79080 & 0.79069 \\
Strict SNR & Strict SNR & 0.79205 & 0.79164 & 0.79206 & 0.79192 \\
Strict SNR & Joint control & 0.78630 & 0.78841 & 0.78721 & 0.78730 \\
\hline
\end{tabular}
\end{adjustbox}
\end{table}

\FloatBarrier
\begin{table}[!ht]
\caption{Full-available dispersion training restores the data-quantity difference at fixed compute. Values are MAE (km s$^{-1}$) for frozen checkpoints after 585 updates and 148,935 path presentations. Matched full and strict training each use 9,929 paths; full available uses all 29,788 eligible fit paths. The three seeds share validation and held-out test sets, architecture, optimization settings, and learning-rate schedule. Lower values indicate better predictions. Means summarize three runs and are not equivalence tests; full per-seed MAE and RMSE are provided in \texttt{data/dispersion\_full\_available\_scores.csv}.}
\label{si:tab:si_dispersion_full_available}
\centering
\small
\setlength{\tabcolsep}{4pt}
\begin{adjustbox}{max width=\linewidth}
\begin{tabular}{llrrrr}
\hline
Test & Training set & Seed 09 & Seed 10 & Seed 11 & Mean \\
\hline
Unfiltered & Matched full & 0.042599 & 0.041282 & 0.041419 & 0.041767 \\
Unfiltered & Strict SNR & 0.040597 & 0.039896 & 0.041319 & 0.040604 \\
Unfiltered & Full available & 0.040809 & 0.039843 & 0.040895 & 0.040516 \\
\addlinespace
Strict SNR & Matched full & 0.043297 & 0.041933 & 0.042463 & 0.042564 \\
Strict SNR & Strict SNR & 0.037259 & 0.036311 & 0.038339 & 0.037303 \\
Strict SNR & Full available & 0.041858 & 0.040500 & 0.041914 & 0.041424 \\
\hline
\end{tabular}
\end{adjustbox}
\end{table}

\end{document}